\documentclass[preprint,superscriptaddress,amsmath,amssymb,aps,floatfix,nofootinbib,longbibliography]{revtex4-1}

\usepackage[pdftex]{graphicx}
\usepackage{bm}
\usepackage{color}
\usepackage{times}
\usepackage{amsmath}
\usepackage{amssymb}
\usepackage{mathtools}
\usepackage{setspace}

\begin{document}

\title{Impact of pressure dissipation on fluid injection into layered aquifers}

\author{Luke T. Jenkins}
\affiliation{Department of Earth Sciences, University of Oxford, Oxford, OX1 3AN, UK}
\affiliation{Department of Engineering Science, University of Oxford, Oxford, OX1 3PJ, UK}

\author{Martino Foschi}
\affiliation{Department of Earth Sciences, University of Oxford, Oxford, OX1 3AN, UK}

\author{Christopher W. MacMinn}
\email{christopher.macminn@eng.ox.ac.uk}
\affiliation{Department of Engineering Science, University of Oxford, Oxford, OX1 3PJ, UK}

\date{\today}


\begin{abstract}
Carbon dioxide (CO$_2$) capture and subsurface storage is one method for reducing anthropogenic CO$_2$ emissions to mitigate climate change. It is well known that large-scale fluid injection into the subsurface leads to a buildup in pressure that gradually spreads and dissipates through lateral and vertical migration of water. This dissipation can have an important feedback on the shape of the CO$_2$ plume during injection, but the impact of vertical pressure dissipation, in particular, remains poorly understood. Here, we investigate the impact of lateral and vertical pressure dissipation on the injection of CO$_2$ into a layered aquifer system. We develop a compressible, two-phase model that couples pressure dissipation to the propagation of a CO$_2$ gravity current. We show that our vertically integrated, sharp-interface model is capable of efficiently and accurately capturing water migration in a layered aquifer system with an arbitrary number of aquifers. We identify two limiting cases --- `no leakage' and `strong leakage' --- in which we derive analytical expressions for the water pressure field for the corresponding single-phase injection problem. We demonstrate that pressure dissipation acts to suppress the formation of an advancing CO$_2$ tongue during injection, reducing the lateral extent of the plume. The properties of the seals and the number of aquifers determine the strength of pressure dissipation and subsequent coupling with the CO$_2$ plume. The impact of pressure dissipation on the shape of the CO$_2$ plume is likely to be important for storage efficiency and security.
\end{abstract}

\maketitle

\section{Introduction}

Carbon capture and geological storage (CCS) involves capturing carbon dioxide (CO$_2$) and injecting it into saline aquifers for long-term storage. The goal of CCS is to reduce CO$_2$ emissions to the atmosphere in order to mitigate climate change~\citep[\textit{e.g.},][]{ipcc-cambridge-2005}. To achieve a meaningful reduction in CO$_2$ emissions, very large quantities of CO$_2$ would need to be captured and stored. Two key physical mechanisms limit the potential storage capacity of a particular aquifer: Pressure buildup and CO$_2$ migration~\citep{szulczewski-pnas-2012}. Pressure buildup limits capacity because the pressure in the target aquifer will increase during injection. The local geology and geomechanics impose a maximum allowable pressure that, if exceeded, could lead to fracturing or fault activation, enabling leakage of CO$_2$ into overlying aquifers. Migration limits capacity because, after injection, the buoyant CO$_2$ will slowly rise, spread, and migrate relative to the denser water. The injection scenario must be designed such that this CO$_2$ will not migrate outside of its designated storage area. Pressure buildup and migration have been studied extensively, but almost exclusively as separate problems due to computational limitations and the widespread view that these processes are essentially independent. Here, we develop a new model that captures both processes simultaneously and we use it to show that they are inherently coupled.

A saline aquifer is a layer of rock with a relatively high permeability, such as sandstone, that is bounded above and below by sealing layers (``seals''), which are layers of rock with much lower permeability, such as shale or mudstone. Aquifers range in thickness from a few metres to a few hundreds of metres; seals are typically about an order of magnitude thinner, from a few centimetres to a few tens of metres. Both aquifers and seals are laterally extensive over tens to hundreds of kilometres, are nearly horizontal (slopes of at most a few degrees), and are saturated with saline groundwater (``water''). A typical sedimentary basin comprises many repetitions of this fundamental sequence (seal-aquifer-seal) over a total thickness of a few kilometres.

Most previous studies of CO$_2$ migration are at the aquifer scale, focusing on the target aquifer only and taking the associated seals to be perfectly impermeable. In this setting, it is common to assume that the CO$_2$ will remain separated from the water by a sharp interface (the capillary pressure being much smaller than the hydrostatic pressure) and that the vertical pressure variation within both fluids will remain essentially hydrostatic (the vertical dimension of the flow being much smaller than the horizontal one). These assumptions together imply that the buoyant CO$_2$ will take the form of a coherent plume known as a gravity current~\citep{huppert-jfm-1995}. The resulting models are convenient for analytical and computational analysis because they eliminate the vertical dimension, leading to a 1D (or 2D) flow problem in the lateral plane. Gravity-current models (also known as ``vertically integrated'' models) have been studied extensively, yielding a variety of important qualitative insights as well as quantitative analytical and semi-analytical predictions~\citep[e.g.,][]{nordbotten2005injection, nordbotten2006similarity, hesse-jfm-2007, gasda-compgeosci-2009, juanes-tpm-2010, Mathias2009approx, macminn2010co2, Pegler2014injection, zheng-jfm-2015, golding-jfm-2017}. However, the majority of these studies provide no insight on pressure buildup or dissipation because they assume that the fluids and the rock are incompressible. Two noteworthy exceptions are the work of \citet{Mathias2009approx, mathias-tipm-2011} and that of \citet{hewitt-jfm-2015}. The former considered the impact of compressibility on CO$_2$ injection, pressure buildup, and lateral pressure dissipation within an isolated aquifer; the latter considered the impact of poroelastic deformation on the same problem, but for a system with incompressible constituents.

The pressure perturbation due to CO$_2$ injection travels orders of magnitude faster and farther than the CO$_2$ itself~\citep[\textit{e.g.},][]{nicot-ijggc-2008, birkholzer2009large, ChangHesse2013}. As a result, most previous studies of pressure buildup consider much larger, basin-scale systems that allow for pressure dissipation via water migration both laterally within aquifers and vertically across seals. Fluid and rock compressibility are central to the rate of pressure buildup and dissipation in these basin-scale systems. Because of the importance of both vertical and lateral flow, models for pressure dissipation are typically fully 2D (or 3D) and are therefore less analytically tractable and more computationally expensive than gravity-current models. The primary computational challenge for these models is resolving the fine-scale features of the long, thin CO$_2$ plume in what is necessarily a large computational domain. As a result, these models typically produce a fairly coarse view of the evolution of the CO$_2$ plume \citep[\textit{e.g.},][]{birkholzer2009large}. Many studies of pressure buildup simplify the problem by replacing CO$_2$ injection with water injection, reducing the two-phase flow problem to a single-phase flow problem. This simplification, which greatly reduces the computational cost, is motivated by the argument that the features of pressure buildup and dissipation away from the target aquifer depend mainly on the rate, duration, and location of injection, but are relatively insensitive to the properties of the injected fluid~\citep{nicot-ijggc-2008, nicot-ghgt-2010, ChangHesse2013}. However, the resulting models not be used to predict anything about the CO$_2$ plume or its coupling with pressure buildup and dissipation.

Studies of pressure buildup and dissipation have consistently shown that vertical pressure dissipation, in particular, has a very strong impact on both overall pressure buildup and lateral pressure propagation~\citep{birkholzer2009large, ChangHesse2013}. This implies that vertical pressure dissipation should also have a strong impact on the shape of the CO$_2$ plume. Here, we show that the shape of the CO$_2$ plume is indeed strongly coupled to vertical pressure dissipation and, further, that this coupling is two-way: Vertical pressure dissipation near the injection well is itself influenced by the shape of the CO$_2$ plume. We do this by developing a novel model that extends the traditional gravity-current approach to allow for compressibility, weak vertical flow, and vertical water migration in a domain comprising an arbitrarily extensive sequence of aquifers and seals. In \S\ref{s:model}, we outline the derivation of the model. The model is computationally inexpensive, but sufficiently complex that analytical solutions are not readily available; in \S\ref{s:results}, we outline our numerical scheme and then benchmark our model against 1D analytical solutions and a full 2D numerical solution for a single-phase model problem (water injection). We then apply our model to CO$_2$ injection and conduct a detailed exploration of the associated parameter space. In \S\ref{s:conclusions}, we conclude by considering the implications our results for CCS.

\section{Theoretical model}\label{s:model}

\begin{figure}
   \centering
  \includegraphics[width=0.5\textwidth]{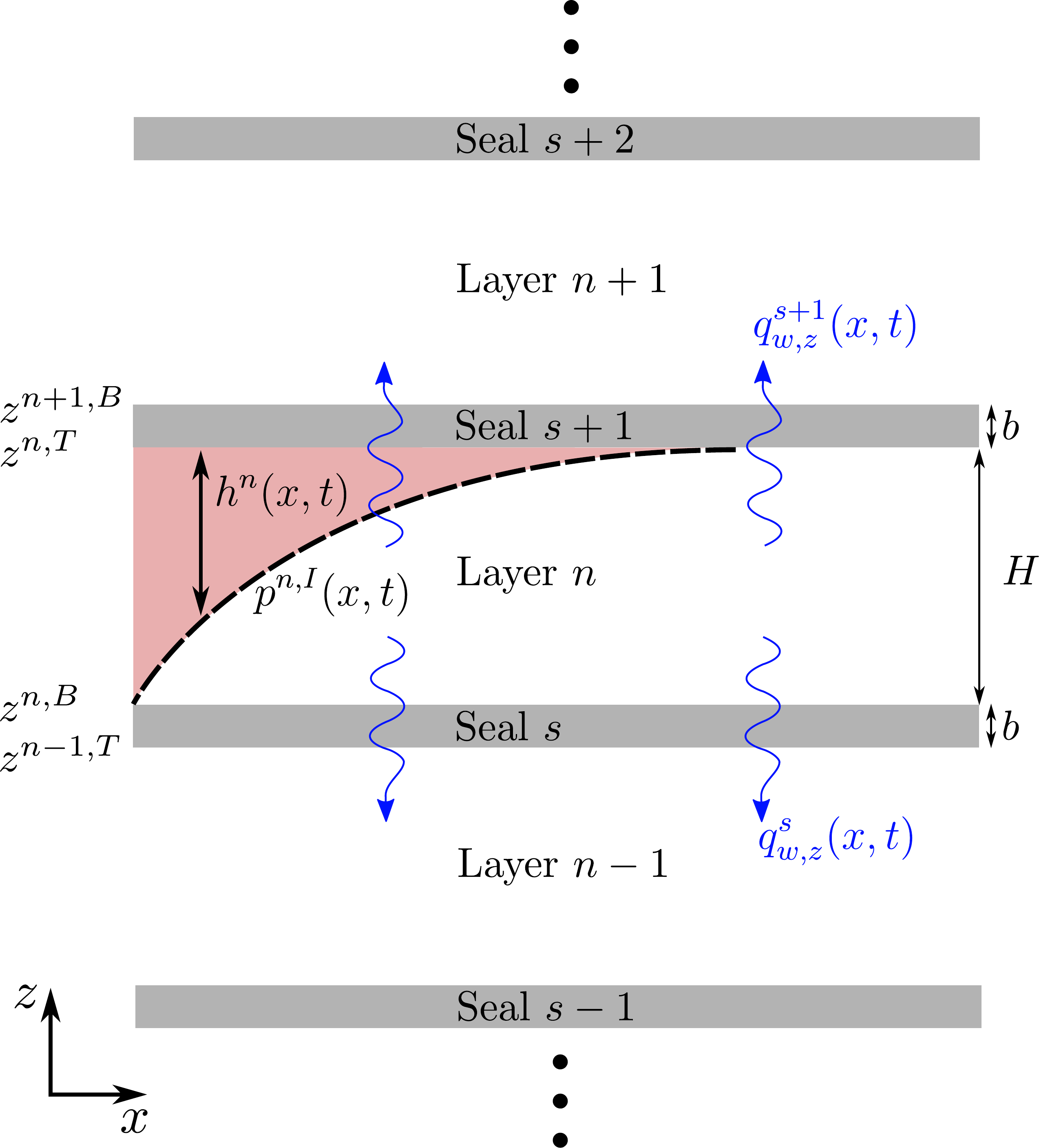}
  \caption{A section of our model system, which comprises a sequence of aquifers of thickness $H$ and seals of thickness $b$. The gas-saturated region is show in red. \label{fig:leak_diagram} }
\end{figure}

The geological setting for our model is a sequence of $N_z$ horizontal aquifers alternating with $N_z+1$ horizontal seals (Figure~\ref{fig:leak_diagram}). For simplicity, we assume that all aquifers have the same uniform thickness $H$, porosity $\phi$, and isotropic permeability $k$, and that all of the thinner and less-permeable intervening seals have the same uniform thickness $b$ ($b\ll{}H$), porosity $\phi_s$, and isotropic permeability $k_s$ ($k_s\ll{}k$). The system is bounded above and below by impermeable seals, and we count aquifers and seals from the bottom up. As a result, the deepest and shallowest seals are seals $1$ and $N_z+1$, respectively, the deepest and shallowest aquifers are aquifers $1$ and $N_z$, respectively, and, in general, aquifer $n$ is bounded by seals $s$ and $s+1$.

In the context of this geological setting, we study the flow of two immiscible phases of different density: A dense, wetting phase and a buoyant, nonwetting phase. The wetting phase is groundwater (``water''); we refer to the buoyant, nonwetting phase as ``gas'' for simplicity, but it could be natural gas, oil, or supercritical CO$_2$.

We denote fluid-phase identity with a subscript $\alpha$, with $\alpha=w$ for water and $\alpha=g$ for gas (or any other buoyant, nonwetting phase). We account for the weak compressibility of both fluids by allowing their densities $\rho_\alpha$ to vary linearly with pressure about a reference state,
\begin{equation}\label{eq:comp}
    \rho_\alpha(p) =\rho_\alpha^0\left[1+c_\alpha(p_\alpha-p^0)\right],
\end{equation}
where $p_\alpha$ is the pressure of phase $\alpha$, $\rho_\alpha^0$ is the density of phase $\alpha$ at reference pressure $p^0$, and $c_\alpha$ is the compressibility of phase $\alpha$ about $p^0$ ($c_\alpha\equiv(1/\rho_{\alpha}^0)(\mathrm{d}\rho_\alpha/\mathrm{d}p)|_{p^0}$).

For the range of pressures typically experienced during both natural fluid migration and subsurface-engineering operations, we expect that $c_w(p_w-p^0)\ll{}1$ and therefore that $\rho_w\approx{}\rho_{w}^{0}$. We take advantage of this simplification in the analysis below. We do not make this assumption for gas since $c_g\gg{}c_w$.

We next develop a model for flow of gas and water in aquifer $n$. To enable vertical and lateral pressure dissipation, we allow for water exchange with the aquifers above and below via flow through the intervening seals. Importantly, we do not allow for gas exchange across the seals (``gas leakage''). For a competent seal, gas leakage is blocked by a large capillary entry pressure due to the fine-grained microstructure of the seal rock. However, gas leakage can be enabled by injection pressures that exceed this entry pressure, by the prescence of heterogeneities in the seal with much lower entry pressure (\textit{e.g.}, sandy patches), or by focused leakage pathways such as faults or fractures. We will address gas leakage in detail in future work.

For simplicity, we focus here on a planar (2D) model problem in the $x$-$z$ plane, with $z$ the vertically upward coordinate and $x$ the horizontal (lateral) coordinate (Figure~\ref{fig:leak_diagram}). We assume symmetry along the $y$ direction (into the page). We denote the vertical position of the top and bottom of aquifer $n$ by $z^{n,T}$ and $z^{n,B}$, respectively, such that $z^{n,T}-z^{n,B}\equiv{}H$.

\subsection{Flow in aquifer $n$} \label{sec:l2}

We begin by assuming that the two fluids are strongly segregated by gravity, such that there exists a region saturated primarily with gas above a region saturated exclusively with water (Figure~\ref{fig:leak_diagram}). The saturation of both fluids would vary in space and time within these regions; however, when capillarity is weak relative to buoyancy, these variations are localised to a relatively thin ``capillary fringe'' that separates a region containing mostly mobile gas from a region containing mostly mobile water, both at nearly constant and uniform saturation within these regions. The evolution of the capillary fringe does not have a leading-order impact on the motion of the gas plume~\citep{Golding2011}. We therefore assume that the gas region and the water region are separated by a sharp interface, and that the gas region contains mobile gas with a uniform and constant saturation of residual water. These assumptions are standard for large-scale gas injection and migration~\citep{nordbotten2005injection, nordbotten2006similarity, hesse-jfm-2007, gasda-compgeosci-2009, juanes-tpm-2010, Mathias2009approx, macminn2010co2}.

As discussed in more detail below, we assume that the residual water exists in a network of connected wetting films and bridges that, although immobile, can conduct a net vertical flow of water. We further assume that the water region in each aquifer contains only water. Residual gas in the water region can be included in a relatively straightforward way~\citep[\textit{e.g.},][]{kochina-intjengsci-1983, barenblatt-cambridge-1996, hesse-jfm-2008, gasda-compgeosci-2009, juanes-tpm-2010, macminn2010co2}, but we neglect it here for simplicity. Lastly, we assume that the seals contain only water.

\subsubsection{Gas in aquifer $n$}\label{sss:conservation_gas}

Conservation of mass for gas in aquifer $n$ is given by
\begin{equation}\label{eq:consG}
    \frac{\partial}{\partial{t}}\left(\rho_g{}\phi{}s_g\right) +\boldsymbol{\nabla}\cdot\left(\rho_g\boldsymbol{q}_g\right) = \mathcal{I}_g,
\end{equation}
where $s_g$ is the saturation of gas, $\boldsymbol{q}_g$ is the Darcy flux of gas, and $\mathcal{I}_g$ is a source term that prescribes the local mass rate of gas injection per unit volume. The Darcy flux of gas is given by Darcy's law,
\begin{equation}\label{eq:DarcyG}
    \boldsymbol{q}_g =-\frac{kk_{rg}}{\mu_g}\,\left(\boldsymbol{\nabla}p_g+\rho_g{}g\hat{\boldsymbol{e}}_z\right),
\end{equation}
where $k_{rg}$ is the relative permeability to gas flow, $\mu_g$ is the dynamic viscosity of gas, which we take to be constant and uniform, $p_g$ is the gas pressure, $g$ is the body force per unit mass due to gravity, and $\hat{\boldsymbol{e}}_z$ is the unit vector in the positive $z$ direction.

We now assume that the gas is in vertical equilibrium, meaning that the vertical component of gas flow is negligible relative to the horizontal components ($q_{g,z}\ll{}q_{g,x}$). This standard assumption is motivated by the long-and-thin aspect ratio typical of these flows, and can be justified rigorously as the leading-order problem under a lubrication-type approximation~\citep{yortsos-tpm-1995, deloubens-jfm-2011}. The vertical pressure distribution in the gas is therefore nearly hydrostatic,
\begin{equation}\label{eq:VFE}
    \frac{\partial{p_g}}{\partial{z}}\approx{}-\rho_{g}g \quad\quad z^{n,I}\leq{}z\leq{}z^{n,T},
\end{equation}
where $z^{n,I}(x,t)$ is the vertical position of the gas-water interface. We then integrate Equation~\eqref{eq:VFE} to arrive at an expression for the vertical pressure distribution in the gas. In doing so, we neglect variations in density due to variations in phase-static pressure over the span of a single aquifer---that is, we assume that $\rho_{g}^{0}gHc_g\ll{}1$. We make use of this assumption repeatedly below. The resulting pressure profile is
\begin{equation}\label{eq:pg_VFE}
    p_g(x,z,t)\approx{}p^n(x,t)-\rho_g^{n}g[z-z^{n,I}(x,t)] \quad\quad z^{n,I}\leq{}z\leq{}z^{n,T},
\end{equation}
where $p^n(x,t)$ is the pressure at the interface, $\rho_g^n(x,t)$ is the vertically averaged gas density, and $h^n(x,t)=z^{n,T}-z^{n,I}(x,t)$ is the thickness of the gas layer. Note that we neglect the capillary pressure at the gas-water interface relative to typical phase-static pressures, $p_c\ll\rho_{g}^{0}gH$, taking the water pressure and the gas pressure to be approximately equal along the interface within each aquifer. This is a standard assumption~\citep[\textit{e.g.,}][]{nordbotten2006improved, hesse-jfm-2007, juanes-tpm-2010}. A constant and uniform capillary pressure can easily be included, but would not change the results below.

Equation~\eqref{eq:pg_VFE} implies that the lateral pressure gradient in the gas is given by
\begin{equation}\label{eq:pg_VFE_dpdx}
    \frac{\partial{p_g}}{\partial{x}}\approx\frac{\partial{p^n}}{\partial{x}}-\rho_g^{n}g\frac{\partial{h^n}}{\partial{x}} \quad\quad z^{n,I}\leq{}z\leq{}z^{n,T},
\end{equation}
where we have again neglected terms of order $\rho_{g}^{0}gHc_g\ll{}1$. Equation~\eqref{eq:pg_VFE_dpdx} implies that the lateral gas flux is given by
\begin{equation}\label{eq:qgx}
    q_{g,x}(x,z,t)\approx-\frac{kk_{rg}}{\mu_g}\left(\frac{\partial{p^n}}{\partial{x}}-\rho_g^{n}g\frac{\partial{h^n}}{\partial{x}}\right) \quad\quad z^{n,I}\leq{}z\leq{}z^{n,T},
\end{equation}
where $k_{rg}$ is the relative permeability to gas in the gas region. Relative permeability is traditionally taken to be a nonlinear and hysteretic constitutive function of saturation, $k_{rg}(s_g)$; however, having assumed that $s_g$ is constant and uniform within the gas region (see beginning of \S{}2.1), we take $k_{rg}$ to be constant and uniform. There is no gas below the interface, so $q_{g,x}=0$ for $z^{n,B}\leq{}z<z^{n,I}$.

We next integrate Equation~\eqref{eq:consG} vertically over the full thickness of aquifer $n$,
\begin{equation}\label{eq:consG_VI}
    \int_{z^{n,B}}^{z^{n,T}}\,\frac{\partial}{\partial{t}}\left(\rho_g{}s_g\phi\right)\,\mathrm{d}z+\int_{z^{n,B}}^{z^{n,T}}\,\boldsymbol{\nabla}\cdot\left(\rho_g\boldsymbol{q}_g\right)\,\mathrm{d}z = \int_{z^{n,B}}^{z^{n,T}}\,\mathcal{I}_g\,\mathrm{d}z.
\end{equation}
This integration procedure is well established~\citep[\textit{e.g.,}][]{Bear1972, gasda-compgeosci-2009}, so we summarise the key results while highlighting the non-standard aspects of our model.

The first term on the left-hand side of Eq.~\eqref{eq:consG_VI} becomes
\begin{equation}\label{eq:G_T1}
    \begin{split}
        \int_{z^{n,B}}^{z^{n,T}}\,\frac{\partial}{\partial{t}}(\rho_g{}\phi{}s_g)\,\mathrm{d}z &\approx{}\frac{\partial}{\partial{t}}(\rho_g^{n}\phi{}s_gh^n) \\
        &\approx\rho_g^{n}\phi{}s_g\bigg[(c_r+  \frac{\rho_{g}^{0}}{\rho_g^{n}}c_g)h^n\frac{\partial{p^n}}{\partial{t}} + \frac{\partial{h^n}}{\partial{t}} \bigg],
    \end{split}
\end{equation}
where $s_g$ is now the constant and uniform saturation of gas in the gas region, $c_r\equiv(1/\phi)(\mathrm{d}{\phi}/\mathrm{d}{p})$ is the rock (matrix) compressibility, and we have again assumed that $\rho_{g}^{0}gHc_g\ll{}1$ \citep{mathias-tipm-2011}. Note that the density ratio multiplying $c_g$ is usually approximated as unity, but this approximation introduces errors in mass conservation of order $c_g(p_g-p^0)$, which is not negligible when the gas is moderately compressible (\textit{e.g.}, in the context of methane migration). Note also that the introduction of rock compressibility to capture the impact of matrix deformation on pressure propagation is a standard and very widely used result from groundwater hydraulics. This approach is strictly valid under the assumptions of negligible lateral strain (\textit{i.e.}, expansion or contraction that is primarily vertical) and constant vertical effective stress (\textit{i.e.}, dominated by gravity), and conveniently decouples pressure propagation from the full machinery of poroelasticity and geomechanics~\citep[\textit{e.g.},][]{vanderkamp-wrr-1983, green-wrr-1990}.

The second term on the left-hand side of Eq.~\eqref{eq:consG_VI} becomes
\begin{equation}\label{eq:G_T2}
    \begin{split}
        \int_{z^{n,B}}^{z^{n,T}}\,\boldsymbol{\nabla}\cdot\left(\rho_g\boldsymbol{q}_g\right)\,\mathrm{d}z &=\frac{\partial}{\partial{x}}\left(\int_{z^{n,B}}^{z^{n,T}}\,\rho_gq_{g,x}\,\mathrm{d}z\right)+(\rho_gq_{g,z})\Big|_{z^{n,B}}^{z^{n,T}} \\
        &\approx\frac{\partial}{\partial{x}}\left[-\rho_g^{n}h^n\frac{kk_{rg}}{\mu_g}\left(\frac{\partial{p^n}}{\partial{x}}-\rho_g^{n}g\frac{\partial{h^n}}{\partial{x}}\right)\right],
    \end{split}
\end{equation}
where we have used Equations~\eqref{eq:DarcyG} and \eqref{eq:pg_VFE_dpdx}, and again assumed that $\rho_{g}^{0}gHc_g\ll{}1$. The vertical gas fluxes at $z=z^{n,B}$ and $z=z^{n,T}$ vanish because we do not allow gas leakage.

Recombining Equations~\eqref{eq:G_T1} and \eqref{eq:G_T2} with Equation~\eqref{eq:consG_VI}, we have
\begin{equation}\label{eq:G_PDE}
    \rho_g^{n}\phi{}s_g\bigg[(c_r+  \frac{\rho_{g}^{0}}{\rho_{g}^{n}}c_g)h^n\frac{\partial{p^n}}{\partial{t}} + \frac{\partial{h^n}}{\partial{t}} \bigg] -\frac{\partial}{\partial{x}}\left[\rho_g^{n}h^n\lambda_g\left(\frac{\partial{p^n}}{\partial{x}}-\rho_g^{n}g\frac{\partial{h^n}}{\partial{x}}\right)\right] = \mathcal{I}_g^{n}H,
\end{equation}
where $\lambda_g\equiv{}kk_{rg}/\mu_g$ is the mobility of gas in the gas region and $\mathcal{I}_g^{n}$ is the vertically averaged mass injection rate of gas per unit volume into aquifer $n$.

\subsubsection{Water in aquifer $n$}

Conservation of mass for the water in aquifer $n$ is given by
\begin{equation}\label{eq:consW}
    \frac{\partial}{\partial{t}}\left(\rho_w{}\phi{}s_w\right) +\boldsymbol{\nabla}\cdot\left(\rho_w\boldsymbol{q}_w\right) = \mathcal{I}_w,
\end{equation}
where $s_w$ is the water saturation, $\boldsymbol{q}_w$ is the Darcy flux of water, and $\mathcal{I}_w$ is a source term that prescribes the local mass rate of water injection per unit volume. The Darcy flux of water is given by Darcy's law,
\begin{equation}\label{eq:DarcyW}
    \boldsymbol{q}_w =-\frac{kk_{rw}}{\mu_w} \,\left(\boldsymbol{\nabla}p_w+\rho_w{}g\hat{\boldsymbol{e}}_z\right),
\end{equation}
where $k_{rw}$ is the relative permeability to water flow, $\mu_w$ is the dynamic viscosity of water, and $p_w$ is the water pressure. The relative permeability to water flow is again typically taken to be a function of water saturation, $k_{rw}(s_w)$; here, our assumptions of no gas in the water region ($s_w=1$ for $z^{n,B}<z<z^{n,I}$) and a constant and uniform saturation of residual water in the gas region ($s_w=1-s_g$ for $z^{n,I}<z<z^{n,T}$) imply that $k_{rw}=1$ in the water region and $k_{rw}=k_{rw}^\star<1$ in the gas region, where $k_{rw}^\star$ is constant and uniform. Note that we also take $\mu_w$ to be constant and uniform.

We assumed above that the gas is in vertical equilibrium, meaning that the vertical component of the gas flux is negligible relative to the horizontal component, such that the vertical pressure distribution is approximately hydrostatic~(Equation~\ref{eq:VFE}). Conversely, for the water we expect a weak but non-negligible vertical flow through the aquifers and the seals. For aquifers containing gas, this upward flow of water must also pass through the connected network of residual water films in the gas region before entering the seal, although the associated relative permeability may be very low (see \S\ref{ss:GI_7L}).

To allow for weak vertical flow of water, we adopt the ansatz that $q_{w,z}$ has the simplest continuous vertical flow structure that allows for different vertical fluxes at the bottom and top of the aquifer: Piecewise linear in $z$. \citet{nordbotten2006improved} suggested this approach in the context of flow near a well in an aquifer with impermeable seals. Here, we extend this approach to account for the gas region and the permeable seals by assuming that $q_{w,z}$ has the following form:
\begin{equation}\label{eq:qwzL2}
    q_{w,z}(x,z,t)\approx
    \begin{cases}
        q_{w,z}^{n,B} + \displaystyle\left(\frac{z-z^{n,B}}{z^{n,I}-z^{n,B}}\right)(q_{w,z}^{n,T}-q_{w,z}^{n,B}) \quad &z^{n,B}\leq{}z<z^{n,I}, \\[1em]
        q_{w,z}^{n,T} \quad &z^{n,I}\leq{}z\leq{}z^{n,T},
    \end{cases}
\end{equation}
such that $q_{w,z}$ in aquifer $n$ varies linearly from to $q_{w,z}^{n,B}(x,t)$ at the bottom seal to $q_{w,z}^{n,T}(x,t)$ at the gas-water interface, and is then uniform and equal to $q_{w,z}^{n,T}(x,t)$ from the gas-water interface to the top seal. This structure implies that water flow in the gas region is primarily vertical, neglecting lateral transport through the gas region relative to lateral transport within the water region. Note that we do not assume anything about the magnitude of the vertical flux or its variation in $x$ or $t$, or about the horizontal flux in the water region---these aspects emerge naturally from Darcy's law and conservation of mass. We discuss the limitations of this assumed structure at the end of \S2.

Equation~\eqref{eq:qwzL2} implies that the vertical pressure variation within the water is given by 
\begin{equation}
    \frac{\partial{p_w}}{\partial{z}}\approx{}
    \begin{cases}
        -\rho_{w}g- \displaystyle\frac{\mu_w}{k}\left[q_{w,z}^{n,B} + \displaystyle\left(\frac{z-z^{n,B}}{z^{n,I}-z^{n,B}}\right)(q_{w,z}^{n,T}-q_{w,z}^{n,B})\right] \quad &z^{n,B}\leq{}z<z^{n,I}, \\[1em]
        -\rho_{w}g- \displaystyle\frac{\mu_w}{kk_{rw}^\star}\,q_{w,z}^{n,T} \quad &z^{n,I}\leq{}z\leq{}z^{n,T}.
    \end{cases}
\end{equation}
We then integrate this expression to arrive at
\begin{equation}\label{eq:pwL2}
    p_w(x,z,t)\approx
    \begin{cases}
        \begin{aligned}
            p^n&+(z^{n,I}-z)\bigg\{\rho_w^{n}g  \\
            &+\displaystyle\frac{\mu_w}{2k}\bigg[(q_{w,z}^{n,T}+q_{w,z}^{n,B})+\left(\frac{z-z^{n,B}}{z^{n,I}-z^{n,B}}\right)(q_{w,z}^{n,T}-q_{w,z}^{n,B})\bigg]\bigg\}
        \end{aligned} \,\, &z^{n,B}\leq{}z<z^{n,I},\\[3em]
        p^n-(z-z^{n,I})\bigg\{\rho_w^{n}g+\displaystyle\frac{\mu_w}{kk_{rw}^\star}\,q_{w,z}^{n,T}\bigg\} \quad &z^{n,I}\leq{}z\leq{}z^{n,T},
    \end{cases}
\end{equation}
where $p^n$ is the pressure along the gas-water interface, which we assume to be the same for both water and gas, as discussed above. As with the derivation for gas, we have neglected variations in density due to variations in hydrostatic pressure over the span of a single aquifer---that is, we have assumed that $\rho_{w}^{0}gHc_w\ll{}1$. Note that $p_w$ is parabolic in $z$ in the water region, linear in $z$ in the gas region, and continuous in $z$ throughout the domain, including across the gas-water interface. We now differentiate Equation~\eqref{eq:pwL2} with respect to $x$ to give the lateral pressure gradient, and thereby the lateral water flux, as presented above for gas. This procedure is straightforward, although more laborious than for gas because $p^n$, $z^{n,I}$, $q_{w,z}^{n,T}$, and $q_{w,z}^{n,B}$ all vary in $x$. The result is
\begin{align}
    \begin{split}
        q_{w,x}(x,z,t)\approx{} -\frac{k}{\mu_w}&\bigg(\frac{\partial{p^n}}{\partial{x}} -\rho_w^ng\frac{\partial{h^n}}{\partial{x}}\bigg) \\
        -\frac{1}{2}\Bigg\{(z^{n,I}-z)&\bigg[\frac{\partial}{\partial{x}}(q_{w,z}^{n,T}+q_{w,z}^{n,B}) +\bigg(\frac{z-z^{n,B}}{z^{n,I}-z^{n,B}}\bigg)\frac{\partial}{\partial{x}}(q_{w,z}^{n,T}-q_{w,z}^{n,B})\bigg] \\
        -\frac{\partial{h^n}}{\partial{x}}&\bigg[(q_{w,z}^{n,T}+q_{w,z}^{n,B})+\bigg(\frac{z-z^{n,B}}{z^{n,I}-z^{n,B}}\bigg)^2(q_{w,z}^{n,T}-q_{w,z}^{n,B})\bigg]\Bigg\}.
    \end{split}
\end{align}
for $z^{n,B}\leq{}z<z^{n,I}$, and recall that we neglect lateral flow of water in the gas region (\textit{i.e.}, $q_{w,x}=0$ for $z^{n,I}\leq{}z\leq{}z^{n,T}$).

Proceeding as above, we next integrate Equation~\eqref{eq:consW} vertically over the full thickness of aquifer $n$,
\begin{equation}\label{eq:consW_VI}
    \int_{z^{n,B}}^{z^{n,T}}\,\frac{\partial}{\partial{t}}\left(\rho_w{}\phi\right)\,\mathrm{d}z+\int_{z^{n,B}}^{z^{n,T}}\,\boldsymbol{\nabla}\cdot\left(\rho_w\boldsymbol{q}_w\right)\,\mathrm{d}z = \int_{z^{n,B}}^{z^{n,T}}\,\mathcal{I}_w\,\mathrm{d}z.
\end{equation}
Much like for gas, the first term on the left-hand side of Equation~\eqref{eq:consW_VI} becomes
\begin{equation}\label{eq:W_T1}
    \begin{split}
        \int_{z^{n,B}}^{z^{n,T}}\,\frac{\partial}{\partial{t}}(\rho_w{}\phi{}s_w)\,\mathrm{d}z &\approx{}\frac{\partial}{\partial{t}}[\rho_w^{n}\phi(H-h^n)+\rho_w^{n}\phi{}(1-s_g)h^n] \\
        &\approx\rho_w^{n}\phi{}\bigg[(c_r+c_w)(H-s_gh^n)\frac{\partial{p^n}}{\partial{t}} - s_g\frac{\partial{h^n}}{\partial{t}} \bigg],
    \end{split}
\end{equation}
where, unlike for gas, we have assumed that $c_w(p_w-p^0)\ll{}1$ and therefore that $\rho_w\approx{}\rho_{w}^{0}$, as discussed above.

The second term on the left-hand side of Equation~\eqref{eq:consW_VI} becomes 
\begin{equation}\label{eq:W_T2}
    \begin{split}
        \int_{z^{n,B}}^{z^{n,T}}\,\boldsymbol{\nabla}\cdot\left(\rho_w\boldsymbol{q}_w\right)\,&\mathrm{d}z =\frac{\partial}{\partial{x}}\bigg(\int_{z^{n,B}}^{z^{n,T}}\,\rho_wq_{w,x}\,\mathrm{d}z\bigg)+(\rho_wq_{w,z})\Big|_{z^{n,B}}^{z^{n,T}} \\
        \approx\frac{\partial}{\partial{x}}\bigg\{-&\rho_{w}^{0}\frac{k}{\mu_w}(H-h^n)\left[\frac{\partial{p^n}}{\partial{x}}-\rho_{w}^{0}g\frac{\partial{h^n}}{\partial{x}}\right] \\
        &-\rho_{w}^{0}\frac{\partial}{\partial{x}}\bigg[\frac{1}{6}(H-h^n)^2(q_{w,z}^{n,B}+2q_{w,z}^{n,T})\bigg]\bigg\}
        +\rho_{w}^{0}(q_{w,z}^{n,T}-q_{w,z}^{n,B}),
    \end{split}
\end{equation}
where we have neglected horizontal flow of water in the gas region ($q_{w,x}\approx{}0$ for $z^{n,I}<z<z^{n,T}$) and again assumed that $\rho_{w}^{0}gHc_w\ll{}1$ and that $\rho_{w}\approx{}\rho_{w}^{0}$. Recombining Equations~\eqref{eq:W_T1} and \eqref{eq:W_T2} with Equation~\eqref{eq:consW_VI}, we have
\begin{equation}\label{eq:W_PDE}
    \begin{split}
        \phi\bigg[(H-s_g{}h^n)(c_r&+c_w)\frac{\partial{p^n}}{\partial{t}}
-s_g\frac{\partial{h^n}}{\partial{t}}\bigg] -\frac{\partial}{\partial{x}}\bigg\{\lambda_w(H-h^n)\bigg[\frac{\partial{p^n}}{\partial{x}} -\rho_{w}g\frac{\partial{h^n}}{\partial{x}}\bigg] \\
    &+\frac{1}{6}\frac{\partial}{\partial{x}}\bigg[(H-h^n)^2(q_{w,z}^{n,B}+2q_{w,z}^{n,T})\bigg]\bigg\} = -(q_{w,z}^{n,T}-q_{w,z}^{n,B}) + \frac{\mathcal{I}_w^{n}H}{\rho_w},
    \end{split}
\end{equation}
where $\lambda_w\equiv{}k/\mu_w$ is the mobility of water in the water region.

Equations~\eqref{eq:G_PDE} and \eqref{eq:W_PDE} are $2N_z$ coupled nonlinear partial differential equations (PDEs) in $p^n$ and $h^n$. For a system with permeable seals, the $N_z$ aquifers are coupled by vertical pressure dissipation and the system is closed via expressions for the vertical water fluxes $q_{w,z}^{n,B}$ and $q_{w,z}^{n,T}$ in terms of $p^n$ and $h^n$.

For a system with impermeable seals, the aquifers are uncoupled and flow and pressurisation are constrained to the injection aquifer. With impermeable seals and no gas, Equation~\eqref{eq:W_PDE} then reduces to the classical groundwater-flow equation from groundwater hydraulics~\citep{Bear1979} (see \S\ref{ss:WI_1L}). With impermeable seals and gas, Equations~\eqref{eq:G_PDE} and \eqref{eq:W_PDE} instead reduce to the widely used model for a gravity current in a horizontal aquifer for an incompressible system \citep[\textit{e.g.},]{Bear1972, huppert-jfm-1995}, and to the model of \citet{Mathias2009approx} for a compressible system (see \S\ref{ss:GI_1L}).

\subsection{Coupling the aquifers with vertical fluxes}

The approach of coupling multiple aquifers with vertical fluxes across the intervening seals was previously suggested by \citet{hunt-wrr-1985}. For incompressible and strictly vertical flow of water through seals, conservation of mass requires that the mass flux of water into seal $s$ from aquifer $n-1$ must equal the mass flux of water out of seal $s$ and into aquifer $n$ at the same position $x$ and time $t$. Taking the water density to be approximately uniform and constant throughout the system ($c_w(p_w-p^0)\ll{}1\implies\rho_w\approx{}\rho_{w}^{0}$), there must then be a single water flux $q_{w,z}^s$ associated with each seal $s$:
\begin{equation}\label{eq:qwz_TBs}
    q_{w,z}^s=q_{w,z}^{n-1,T}=q_{w,z}^{n,B}.
\end{equation}
We calculate this flux via Darcy's law,
\begin{equation}\label{eq:qwz_s}
    q_{w,z}^s=-\frac{k_s}{\mu_w}\left(\frac{p_w^{n,B}-p_w^{n-
    1,T}}{b}+\rho_w{}g\right),
\end{equation}
where our assumption of no gas in the seals implies that $k_{rw}=1$, and where $p_w^{n,B}=p_w(z^{n,B})$ and $p_w^{n-1,T}=p_w(z^{n-1,T})$. We write these unknown pressures in terms of the fluxes through the seals by combining Equation~\eqref{eq:pwL2} with Equation~\eqref{eq:qwz_TBs},
\begin{subequations}\label{eq:pwn_TB}
    \begin{align}
        p_w^{n-1,T}&=p^{n-1}-h^{n-1}\bigg[\rho_wg+\frac{\mu_w}{kk_{rw}}q_{w,z}^s\bigg], \\
        p_w^{n,B}&=p^n+(H-h^n)\bigg[\rho_wg+\frac{\mu_w}{2k}(q_{w,z}^{s+1}+q_{w,z}^s)\bigg].
    \end{align}
\end{subequations}
Combining Equations~\eqref{eq:qwz_s} and \eqref{eq:pwn_TB} and rearranging, we arrive at
\begin{equation}\label{eq:qs_system}
    \begin{split}
        \left(\frac{H-h^n}{2\lambda_w}\right)q_{w,z}^{s+1} +\bigg(\frac{h^{n-1}}{\lambda_w^\star}+\frac{b}{\lambda_w^s}&+\frac{H-h^n}{2\lambda_w}\bigg)q_{w,z}^s \\
        &=-\bigg[p^n-p^{n-1}+\rho_{w}^{0}g(h^{n-1}+b+H-h^n)\bigg],
    \end{split}
\end{equation}
where $\lambda_w^\star=kk_{rw}^\star/\mu_w$ is the mobility of water in the gas regions of the aquifers and $\lambda_s=k_s/\mu_w$ is the mobility of water in the seals. Equation~\eqref{eq:qs_system} is a linear system of $N_z-1$ coupled algebraic equations in the $N_z-1$ unknown fluxes $q_{w,z}^s$ for $s=2\ldots{}N_z$, from which we can solve for $q_{w,z}^s$ in terms of $p^n$ and $h^n$. Recall that the bottom-most and top-most seals are impermeable, so that $q_{w,z}^1=q_{w,z}^{N_z+1}=0$. 

\subsection{Boundary and initial conditions}

We consider a system comprised of $N_z$ aquifers and $N_z+1$ seals, and which extends horizontally from $x=-L_x/2$ to $x=L_x/2$. We assume that the system is initially fully saturated with water (no gas), and that the initial pressure distribution is hydrostatic. For simplicity, we also assume that the pressure at the lateral boundaries remains hydrostatic throughout; this implies that our results are independent of lateral domain size for scenarios where the system is sufficiently laterally extensive that changes in pressure due to injection never reach the boundaries, which is true for our reference case (see Figure~\ref{fig:WI_1L_pressures}).

For injection of phase $\alpha$ into the horizontal centre of aquifer $n$ at a mass flow rate $\dot{M}_\alpha^n(t)$ per unit length into the page, the relevant vertically integrated source term $\mathcal{I}_\alpha^n$ can be written
\begin{equation}
    \mathcal{I}_\alpha^n = \frac{\dot{M}_\alpha^n(t)}{H}\,\delta(x),
\end{equation}
where $\delta(x)$ is the Dirac delta function.

\subsection{Non-dimensionalization}

We consider the injection of gas at a mass flow rate $\dot{M}$ per unit length into the page for a time $\mathcal{T}$. This scenario motivates the following characteristic scales for length, pressure, and vertical flux:
\begin{equation}\label{eq:scales}
    \mathcal{L} \equiv \frac{\dot{M}\mathcal{T}}{2\phi s_g \rho_{g}^{0}H}\,\,, \quad\mathcal{P} \equiv\frac{\phi\mathcal{L}^2}{\lambda_w\mathcal{T}}=\frac{\dot{M}\mathcal{L}}{2\lambda_w s_g \rho_{g}^{0}H}\,\,, \quad\mathrm{and}\quad \mathcal{Q}_z \equiv\frac{\lambda_w^s \mathcal{P}}{b}.
\end{equation}
The characteristic length $\mathcal{L}$ is the half-width of an incompressible plug (box) of gas of mass $\dot{M}\mathcal{T}$ per unit length into the page. The characteristic pressure $\mathcal{P}$ is the pressure drop associated with a Darcy flux $\phi\mathcal{L}/\mathcal{T}$ of water over a distance $\mathcal{L}$. The characteristic vertical flux $\mathcal{Q}_z$ is the vertical flux of water associated with a characteristic pressure drop $\mathcal{P}$ across a seal of thickness $b$.

We use the above scales in combination with existing parameters to define the following dimensionless quantities:
\begin{equation}
    \begin{split}
        \tilde{x}\equiv\frac{x}{\mathcal{L}}, \quad
        \tilde{t}\equiv\frac{t}{\mathcal{T}}, \quad
        &\tilde{h}\equiv\frac{h}{H}, \quad \tilde{p}\equiv\frac{p}{\mathcal{P}}, \quad \tilde{q}\equiv\frac{q}{\mathcal{Q}_z}, \\
        &\tilde{b}\equiv\frac{b}{H}, \quad
        \tilde{\rho}_\alpha\equiv\frac{\rho_\alpha}{ \rho_g^0}, \quad
        \tilde{\mathcal{I}}_{\alpha}^{n}\equiv\frac{2\mathcal{L}H\,\mathcal{I}_{\alpha}^{n}}{\dot{M}}.
    \end{split}
\end{equation}
We then also introduce the following dimensionless groups:
\begin{subequations}
    \begin{align}
        N_{cw} &\equiv c_w \mathcal{P}\\
        R_{cw} &\equiv c_r/c_w\\
        R_{cf} &\equiv c_g/c_w\\
        R_A &\equiv \mathcal{L}/H\\
        R_d &\equiv \rho_{g}^{0}/\rho_{w}^{0}\\
        N_g &\equiv \rho_{w}^{0}gH/\mathcal{P}\\
        \mathcal{M} &\equiv \lambda_g/(s_g\lambda_w)\\
        \Lambda_w^s &\equiv \lambda_w^sH/(\lambda_w b)
    \end{align}
\end{subequations}
The first three of these groups capture the effects of the compressibility: The `compressibility number' $N_{cw}$ measures the overall importance of compressibility within the system, whereas the two compressibility ratios $R_{cw}$ and $R_{cf}$ compare the compressibilities of the various phases. The aspect ratio $R_A$ compares the characteristic length of the plume to the aquifer thickness, capturing the importance of horizontal-to-vertical flow within and around the gas plume. The density ratio $R_d$ compares the fluid densities and the `gravity number' compares hydrostatic pressure to the characteristic injection pressure, such that the grouping $(1-R_d)N_g$ measures the importance of buoyancy relative to injection. The mobility ratio $\mathcal{M}$ compares the mobility of gas within the aquifer to that of water, incorporating the gas saturation for convenience. Lastly, the `leakage number' $\Lambda_w^s$ measures the resistance to vertical flow through the aquifers relative to the seals, such that the grouping $R_A^2\Lambda_w^s$ measures the importance of vertical pressure dissipation relative to lateral pressure dissipation.

\subsection{Model summary}

Dropping the tildes, we can now write our coupled partial differential system in dimensionless form as:
\begin{equation}\label{eq:GOV_NDg}
    N_{cw}(R_{cw}\rho_{g}^{n}+R_{cf}) h^n\frac{\partial{p^n}}{\partial{t}} 
+ \rho_{g}^{n}\frac{\partial{h^n}}{\partial{t}} -\mathcal{M}\frac{\partial}{\partial{x}}\bigg[\rho_{g}^{n} h^n\bigg(\frac{\partial{p^n}}{\partial{x}} 
-\rho_{g}^{n}R_dN_g\frac{\partial{h^n}}{\partial{x}}\bigg)\bigg] = \mathcal{I}_{g}^{n},
\end{equation}
and
\begin{equation}\label{eq:GOV_NDw}
    \begin{split}
        N_{cw}(&R_{cw}+1)(1-s_gh^n)\frac{\partial{p^n}}{\partial{t}}
-s_g\frac{\partial{h^n}}{\partial{t}}
-\frac{\partial}{\partial{x}}\bigg\{(1-h^n)\bigg[\frac{\partial{p^n}}{\partial{x}} 
-N_g\frac{\partial{h^n}}{\partial{x}}\bigg] \\
&+\frac{\Lambda_w^s}{6}\frac{\partial}{\partial{x}}\bigg[(1-h^n)^2(q_{w,z}^{s}+2q_{w,z}^{s+1})\bigg]\bigg\}
= -{R_A}^2\Lambda_w^s(q_{w,z}^{s+1}-q_{w,z}^{s}) + s_gR_d\mathcal{I}_w^n
    \end{split}
\end{equation}
with
\begin{equation}\label{eq:dens_ND}
    \rho_g^n(p^n) = 1 + N_{cw}R_{cf}(p^n-p^0)
\end{equation}
and
\begin{equation}\label{eq:qs_system_ND}
    \begin{split}
        \frac{\Lambda_w^s}{2}(1-h^n)q_{w,z}^{s+1} +\bigg[\frac{\Lambda_w^s}{k_{rw}^\star}h^{n-1}+1+&\frac{\Lambda_w^s}{2}(1-h^n)\bigg]q_{w,z}^s \\
        &=-\bigg[p^n-p^{n-1}+N_g(h^{n-1}+b+1-h^n)\bigg].
    \end{split}
\end{equation}
For a system with $N_z$ aquifers, Equations~\eqref{eq:GOV_NDg} and~\eqref{eq:GOV_NDw} provide $2N_z$ coupled PDEs enforcing conservation of mass for gas and for water, respectively, in each aquifer $n=1\ldots{}N_z$. Equation~\eqref{eq:dens_ND} is the dimensionless form of the linear constitutive relationship for gas density. Lastly, Equation~\eqref{eq:qs_system_ND} is a linear system of $N_z-1$ algebraic equations in the dimensionless vertical fluxes of water across each interior seal, $q_{w,z}^s$ for $s=2\ldots{}N_z$, where we impose $q_{w,z}^1=q_{w,z}^{N_z+1}=0$.

We assume that there is initially no gas in the system, $h^n(x,t=0)=0$, and we ensure that the gas never reaches the lateral boundaries. We assume that the pressure is initially hydrostatic, and that the pressure at the boundaries remains hydrostatic:
\begin{equation}\label{eq:pBCs_ICs}
    p^n(x,t=0)=p^n(-L_x/2,t)=p^n(L_x/2,t)=p^0 -N_g[n + (n-1)b],
\end{equation}
recalling that the pressures $p^n$ are the pressures at the gas-water interface in each aquifer, which is the top of the aquifer in the absence of gas, and that $p^0$ is the initial pressure at the bottom of aquifer~1.

Vertical pressure dissipation does not appear explicitly in Equation~\eqref{eq:GOV_NDg} because the gas is limited to vertical equilibrium, which is valid for $R_A\gg{}1$. Vertical pressure dissipation is responsible for two of the terms in Equation~\eqref{eq:GOV_NDw}, both of which are multiplied by $\Lambda_w^s$. The term on the right-hand side measures the net mass of water that enters layer $n$ through seals $s$ and $s+1$, and its dimensionless coefficient measures the importance vertical pressure dissipation relative to lateral pressure dissipation: A pressure difference of size $\mathcal{P}$ over a lateral distance $\mathcal{L}$ would drive a flow rate $Q_l\sim{}\lambda_w\mathcal{P}H/\mathcal{L}$ laterally through the aquifer, and a flow rate $Q_s\sim{}\lambda_w^s\mathcal{P}\mathcal{L}/b$ vertically across the associated seals. The ratio of these flow rates is $Q_s/Q_l=\lambda_w^s\mathcal{L}^2/(\lambda_wHb)=R_A^2\Lambda_w^s$, highlighting that the extensive contact area between the aquifers and the seals enables vertical pressure dissipation to have a strong impact on the pressure field even when $\lambda_w^s\ll{}\lambda_w$ ($\Lambda_w^s\ll{}1$). The term proportional to $\Lambda_w^s$ on the left-hand side is a consequence of conservation of mass, introducing weak lateral variations in $q_{w,x}$ to compensate for vertical variations in $q_{w,z}$. Our assumed piecewise-linear structure for the vertical flux determines the specific structure of this term, but it will always be proportional to $\Lambda_w^s$ and involve the horizontal divergence of some function of $h^n$ times the fluxes through the seals. Our model is valid as long as this term is indeed a weak perturbation to horizontal flow, meaning that $\Lambda_w^s\ll{}1$. In general, this term is clearly less important than the one proportional to $R_A^2\Lambda_w^s$ since $R_A\gg{}1$, and could safely be neglected, but we retain it to preserve the consistency of our formulation.

\section{Results}\label{s:results}

For illustrative purposes, we consider a reference scenario involving fluid injection into the central aquifer ($n=4$) of a seven-aquifer system ($N_z=7$). We choose rock properties consistent with sandstone aquifers and mudstone seals and we choose fluid properties consistent with water and CO$_2$ at a depth of $\sim$1~km, where our reference pressure is the pressure at the bottom of aquifer~1. We consider an injection rate of $\sim$1~Mt per year distributed along a 30~km long array of injection wells for a period of 10~years~\citep{szulczewski-pnas-2012}. Based on this scenario, we choose a set of reference values for our dimensional parameters and then use these to calculate corresponding reference values for our dimensionless parameters. Both sets of values are reported in table~\ref{tab:params}. We use these values in the rest of this study except where noted otherwise.

\begin{table}
    \begin{center}
        \begin{tabular}{ l c l }
            \textbf{Parameter} & \textbf{Symbol} & \textbf{Value} \\
            \\
            Number of aquifers & $N_z$ & 7 \\
            Horizontal extent & $L_x$ & 100 km \\
            Aquifer thickness & $H$ & 10 m \\
            Aquifer porosity & $\phi$ & 0.3 \\
            Aquifer permeability & $k$ & $10^{-13}$ m$^2$ \\
            Seal thickness & $b$ & 1 m \\
            Seal permeability & $k_s$ & $10^{-18}$ m$^2$ \\
            Rock compressibility & $c_r$ & 3.0 $\times 10^{-11}$ Pa$^{-1} $ \\
            Reference pressure & $p^0$ & 10 MPa \\
            \\
            Water viscosity & $\mu_w$ & 8 $\times 10^{-4}$ Pa$\cdot$s \\
            Water density & $\rho_w$ & 1000 kg$\cdot$m$^{-3}$ \\
            Water compressibility & $c_w$ & 4.5 $\times 10^{-10}$ Pa$^{-1}$ \\
            Saturation of water in gas region & $s_{wr}$ & 0.2 \\
            Relative permeability to water in gas region & $k_{rw}^\star$ & 0.01 \\
            \\
            Gas viscosity & $\mu_g$ & 4 $\times 10^{-5}$ Pa$\cdot$s \\
            Gas density & $\rho_{g}^{0}$ & 700 kg$\cdot$m$^{-3}$ \\
            Gas compressibility & $c_g$ & 1.5 $\times 10^{-8}$ Pa$^{-1}$ \\
            Saturation of gas in gas region & $s_g$ & 0.8 \\
            Relative permeability to gas in gas region & $k_{rg}$ & 1 \\
            \\
            Mass injection rate & $\dot{M}$ & $10^{-3}$~kg$\cdot$s$^{-1}$$\cdot$m$^{-1}$ \\
            Injection time & $\mathcal{T}$ & 10 years \\
            \hline\hline
            Compressibility number & $N_{cw}$ & 3.02$\times 10^{-5}$ \\
            Rock-to-water compressibility ratio & $R_{cw}$ & 6.67$\times 10^{-2}$ \\
            Gas-to-water compressibility ratio & $R_{cf}$ & 33.3 \\
            Aspect ratio & $R_A$ & 9.39 \\
            Density ratio & $R_d$ & 0.7 \\
            Gravity number & $N_g$ & 1.46 \\
            Mobility ratio & $\mathcal{M}$ & 25 \\
            Water-leakage strength & $\Lambda_w^s$ & $10^{-4}$ \\
            Seal-to-aquifer thickness ratio & $\tilde{b}$ & 0.1 \\
            Horizontal extent & $\tilde{L}_x$ & 1062 \\
        \end{tabular}
    \end{center}
    \caption{Reference parameter values. The dimensionless values (below the double-line) are calculated directly from the dimensional values (above the double-line). \label{tab:params} }
\end{table}

In the context of this reference scenario, we consider the predictions of our model for several test problems: (i)~water injection with impermeable seals, which allows us to verify our model against a classical analytical solution; (ii)~water injection with permeable seals, which allows us to benchmark our model against a fully 2D groundwater-flow model; (iii)~gas injection with impermeable seals, which allows us to verify our model against previous results for gas injection; and (iv)~gas injection with permeable seals, which allows us to study the impact of pressure dissipation on gas injection.

In all cases, we solve our model numerically by discretising in space using a standard finite-volume method on a uniform grid and then integrating in time using \verb+MATLAB+'s built-in adaptive implicit solver for stiff ODEs, \verb+ODE15s+ \citep[][]{matlab}. In cases with permeable seals, the linear system of equations for the leakage fluxes (Equation~\ref{eq:qs_system_ND}) becomes $N_x$ uncoupled linear algebraic systems of size $N_z-1$, where $N_x$ is the number of horizontal gridblocks; we invert these systems at each timestep using a standard linear solver, which is computationally inexpensive.

\subsection{Water injection with impermeable seals}\label{ss:WI_1L}

We first consider water injection into a one-aquifer system ($n=N_z=1$) with impermeable seals ($\Lambda_w^s=0$) and containing no gas ($h^1=0$), in which case our model reduces to the classical linear groundwater-flow equation from hydrology and hydrogeology,
\begin{equation}\label{eq:lin_diff_one-layer}
    N_{cw}(R_{cw}+1)\frac{\partial{p^1}}{\partial{t}} -\frac{\partial^2{p^1}}{\partial{x^2}} =s_gR_d \mathcal{I}_w^1(x)=2s_gR_d\,\delta(x)\,u(t),
\end{equation}
where we have taken $\dot{M}_g^1=0$ and $\dot{M}_w^1=\dot{M}u(t)$, where $u(t)$ is the unit (Heaviside) step function and $\delta(x)$ is now the dimensionless Dirac delta function. The factor of $s_gR_d$ on the right-hand side is an artefact of our use of a characteristic length based on gas injection (Equation~\ref{eq:scales}). It is awkward for gas properties to appear in a problem with no gas, and they could be eliminated by suitable rescaling of the characteristic length, but their values have no impact on the dimensional solution.

The pressure $p^1$ is the pressure along the gas-water interface in the aquifer (see \S\ref{sss:conservation_gas}). In the absence of gas, this degenerates to the pressure at the top of the aquifer. Recall that our reference pressure is the initial pressure at the bottom of aquifer~1; in the absence of vertical flow, the dimensionless pressure at the top will be lower than the dimensionless pressure at the bottom by the dimensionless hydrostatic contribution over the thickness of one aquifer, $N_g$. We therefore impose the following initial and boundary conditions: $p^1(x,0)=p^1(-L_x/2,t)=p^1(L_x/2,t)=p^0-N_g$.

Equation~\eqref{eq:lin_diff_one-layer} is a linear diffusion problem that can be solved analytically. To do so, we assume symmetry across $x=0$ and focus on the positive sub-domain $0\leq{}x\leq{}L_x/2$. We then rewrite the injection term as a boundary condition: $\partial{p^1}/\partial{x}(0,t)=-s_gR_du(t)$. Standard separation of variables then yields
\begin{equation}
    p^1(x,t)=
    \begin{cases}
        p^0-N_g + s_gR_d(x+L_x/2) + \Omega (x,t) & \text{for} \,\, x\leq0, \\[1em]
        p^0-N_g - s_gR_d(x-L_x/2) + \Omega (x,t) & \text{for} \,\, x\geq0,
    \end{cases}
\end{equation}
where $\Omega(x,t)$ is given by
\begin{equation}\label{eq:Omega}
    \Omega(x,t)=-\sum_{n=0}^{\infty} \frac{4s_gR_d}{L_x\lambda_n^2} \exp\left[-\frac{\lambda_n^2 t}{N_{cw}(R_{cw}+1)}\right]\cos(\lambda_n x)
\end{equation}
with
\begin{equation}
    \lambda_n = \frac{(2n+1)\pi}{L_x}.
\end{equation}
This solution is well known, and it is not surprising that our numerical scheme can reproduce it. We use it here as an instructive reminder of the impact of compressibility on lateral pressure dissipation.

\begin{figure}
    \centering
    \includegraphics[width=\textwidth]{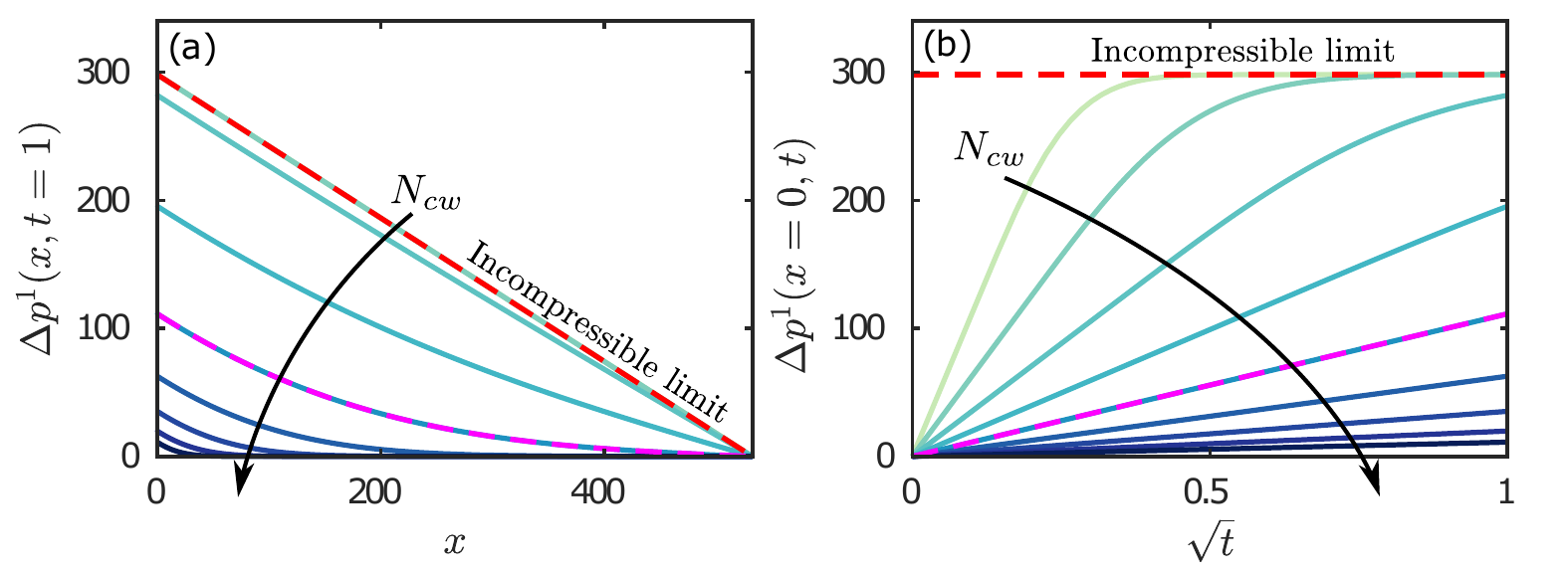}
    \caption{Pressure perturbation during water injection into a one-aquifer system with impermeable seals for a wide range of compressibilities, $\log_{10}(N_{cw})\approx{}-6.5$, $-6$, $-5.5$, $-5$, $-4.5$, $-4$, $-3.5$, $-3$, and $-2.5$. We plot (a)~the pressure perturbation at the end of injection, $\Delta{p^1}(x,t=1)$, against $x$, and (b)~the injection pressure, $\Delta{p^1}(x=0,t)$, against $\sqrt{t}$. The dashed red line is the incompressible limit ($N_{cw}\to{}0$) and the dashed magenta line is the analytical solution for the reference scenario ($\log_{10}(N_{cw})\approx{}-4.5$). Note that the curve for $\log_{10}(N_{cw})\approx{}-6.5$ is not visible in panel (a). \label{fig:WI_1L_pressures} }
\end{figure}

In Figure~\ref{fig:WI_1L_pressures}, we plot the pressure perturbation due to injection, $\Delta{p^1}\equiv{}p^1-(p^0-N_g)$, for different values of the compressibility number $N_{cw}$.
Recall that $N_{cw}$ compares the characteristic compressibility to the characteristic pressure, so that larger values (due either to larger compressibility or to larger pressure) imply an increasingly compressible system, whereas $N_{cw}\to{}0$ implies an incompressible system. As $N_{cw}$ increases, $\Delta{p^1}$ is smaller and more concentrated near the injection point ($x=0$)---that is, compressibility mitigates pressure buildup in both magnitude and extent for a fixed injection time (Figure~\ref{fig:WI_1L_pressures}a). For all values of $N_{cw}$, the injection pressure is proportional to $\sqrt{t}$ until the perturbation reaches the boundaries (Figure~\ref{fig:WI_1L_pressures}b). Thereafter, interaction with the fixed pressure at $x=\pm{}L_x$ drives a transition toward a steady-state profile that is linear in $x$. The system reaches this steady state more quickly as $N_{cw}$ decreases, and instantaneously in the incompressible limit ($N_{cw}\to{}0$).

\subsection{Water injection with permeable seals}\label{ss:WI_7L}

Our model captures vertical pressure dissipation at the basin scale by allowing for a weak vertical flow of water through the aquifers and across the seals, assuming that this vertical water flux has a continuous, piecewise-linear structure. We also neglect compressibility and lateral transport within the seals by assuming that the vertical fluxes of water in and out must be equal.

To test these assumptions, we next consider water injection into the central aquifer ($n=4$) of a seven-aquifer system ($N_z=7$) with permeable seals, but no gas ($h^n=0$ for all $n$). In this context, our model is most similar to the linear groundwater-flow model proposed by \citet{hunt-wrr-1985} for coupling layered aquifers via vertical flow across the intervening seals; \citet{hunt-wrr-1985} neglected vertical flow except across the seals, whereas we allow for weak vertical flow throughout the system. Our model becomes
\begin{equation}\label{eq:lin_diff_seven-layer}
    \begin{split}
        N_{cw}(R_{cw}+1)\frac{\partial{p^n}}{\partial{t}} -\frac{\partial}{\partial{x}}\bigg[\frac{\partial{p^n}}{\partial{x}}+\frac{\Lambda_w^s}{6}\frac{\partial}{\partial{x}}\bigg(&q_{w,z}^s+2q_{w,z}^{s+1}\bigg)\bigg] \\
        &=-R_A^2\Lambda_w^s(q_{w,z}^{s+1}-q_{w,z}^s)+s_gR_d \mathcal{I}_w^n(x),
    \end{split}
\end{equation}
for $n=s=1\ldots{}7$, with
\begin{equation}
    \frac{\Lambda_w^s}{2}q_{w,z}^{s+1} +\bigg[1+\frac{\Lambda_w^s}{2}\bigg]q_{w,z}^s =-\bigg[p^n-p^{n-1}+N_g(b+1)\bigg],
\end{equation}
for $n=s=2\ldots{}6$, where $\mathcal{I}_w^4=2s_gR_d\delta(x)u(t)$, $\mathcal{I}_w^n=0$ for $n\neq{}4$, and $q_{w,z}^1=q_{w,z}^8=0$. The initial and boundary conditions are as in Equation~\eqref{eq:pBCs_ICs}.

To assess the accuracy of our model, we compare it with a classical 2D groundwater-flow model, which can be written in our notation as
\begin{equation}\label{eq:lin_diff_seven-layer_2D}
    N_{cw}(R_{cw}+1)\frac{\partial{p}}{\partial{t}} -\Lambda_w^{\mathrm{2D}}\left(\frac{\partial^2{p}}{\partial{x^2}}+\frac{\partial^2{p}}{\partial{z^2}}\right) =s_gR_d\mathcal{I}_w^{\mathrm{2D}},
\end{equation}
where $p(x,z,t)$ is the full 2D pressure field, $\Lambda_w^{\mathrm{2D}}$ is equal to 1 within the aquifers and to $b\Lambda_w^s$ within the seals, and $\mathcal{I}_w^{\mathrm{2D}}$ is equal to $2s_gR_d\delta(x)u(t)$ within aquifer~4 and to 0 elsewhere. This model allows for full 2D flow, as well as compressibility in both the aquifers and the seals~\citep[\textit{e.g.},][]{Bear1979}. We impose the following initial and boundary conditions: $p(x,z,0)=p(-L_x/2,z,t)=p(L_x/2,z,t)=p^0-N_g{}(z-z^{1,B})$ and $\partial{p}/\partial{z}(x,z^{1,B},t)=\partial{p}/\partial{z}(x,z^{7,T},t)=0$. As with our reduced-order model, we solve Equation~\eqref{eq:lin_diff_seven-layer_2D} numerically by discretising in space using a standard finite-volume method on a uniform 2D grid and integrating in time using \verb+MATLAB+'s built-in implicit solver for stiff ODEs, \verb+ODE15s+ \citep{matlab}. To ensure a fair comparison between the two models, we use the same resolution in $x$ and the same absolute and relative tolerances in time for both solutions.

\begin{figure}
    \centering
   \includegraphics[width=\textwidth]{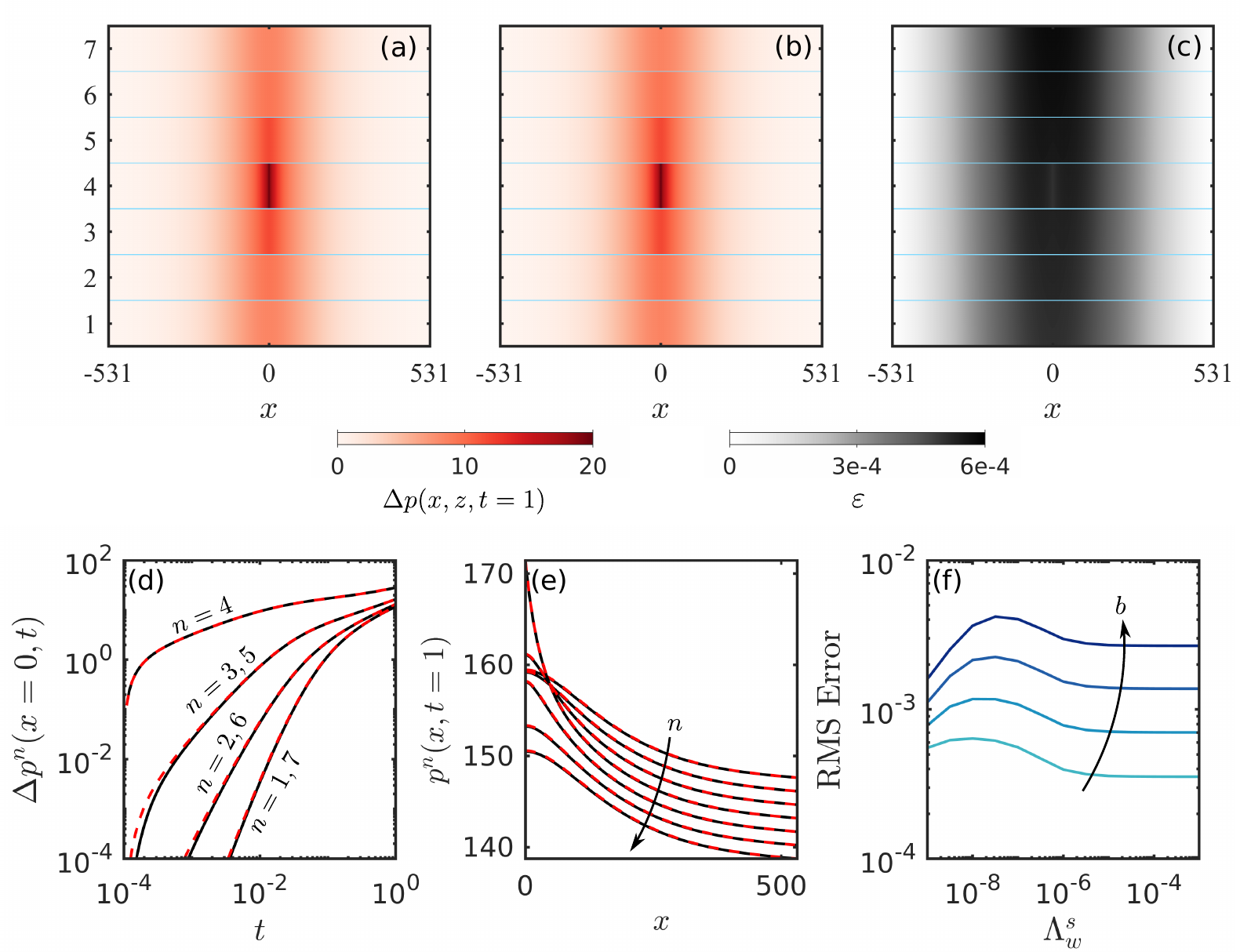}
    \caption{Pressure perturbation during water injection into aquifer~4 of a seven-aquifer system with permeable seals, comparing our quasi-2D model against a full 2D model. We show (a)~the reconstructed 2D pressure perturbation from our quasi-2D model at $t=1$ for $b=0.0125$ and $\Lambda_w^s=10^{-5}$, (b)~the same thing for the full 2D model, and (c)~the magnitude of the relative error $\varepsilon$ between the two, disregarding the seals (shown in light blue). We also plot (d)~the pressure perturbation at the top of each aquifer at $x=0$ against $t$ for our quasi-2D model (dashed red) and the full 2D model (solid black), (e)~the pressure at the top of each aquifer at $t=1$ against $x$ for the quasi-2D model (dashed red) and the full 2D model (solid black), and (f) the root-mean-square (RMS) relative error between the two models at $t=1$ against $\Lambda_w^s$ for $b=0.0125$, $0.025$, $0.05$, and $0.1$. \label{fig:WI_7L_benchmark} }
\end{figure}

In Figure~\ref{fig:WI_7L_benchmark}(a), we show the pressure perturbation predicted by our quasi-2D model at the end of injection ($t=1$) for $b=0.0125$ and $\Lambda_w^s=10^{-5}$. We show the same quantity for the full 2D model in Figure~\ref{fig:WI_7L_benchmark}(b), and the relative error between the two in Figure~\ref{fig:WI_7L_benchmark}(c). For reference, solving the full 2D model took about 8~minutes on a single core of a standard desktop PC, whereas solving our quasi-2D model for the same scenario took about 0.4~seconds. We compare these predictions in more detail in Figures~\ref{fig:WI_7L_benchmark}(d), \ref{fig:WI_7L_benchmark}(e), and \ref{fig:WI_7L_benchmark}(f). Because our model neglects compressibility within the seals, it begins pressurising the overlying and underlying aquifers slightly faster than the full 2D model, leading to a small disagreement in $p^n$ outside of the injection layer at early times~(Figure~\ref{fig:WI_7L_benchmark}d). This error decays as the seals pressurise, which happens over a dimensionless timescale $bN_{cw}(1+R_{cw})/(R_A^2\Lambda_w^s)\ll1$ (dimensional timescale $b^2\phi(c_w+c_r)/\lambda_w^s\ll{}\mathcal{T}$). For the scenario shown in Figure~\ref{fig:WI_7L_benchmark}(a--e), this timescale is $bN_{cw}(1+R_{cw})/(R_A^2\Lambda_w^s)\approx{}4\times{}10^{-4}$ and the error does indeed become negligible for dimensionless times sufficiently greater than this value. Figure~\ref{fig:WI_7L_benchmark}(f) shows that this source of error increases monotonically with $b$, but varies non-monotically with $\Lambda_w^s$: For small values of $\Lambda_w^s$, the seals pressurise more slowly but vertical pressure dissipation is less important, whereas for larger values of $\Lambda_w^s$, vertical pressure dissipation is more important and the seals pressurise more quickly. The maximum root-mean-square (RMS) relative error between the two solutions is about $0.004$ for $b=0.1$, the largest value tested. These results suggest that our 1D model reproduces the full 2D pressure field both accurately and efficiently.

\begin{figure}
    \centering
    \includegraphics[width=\textwidth]{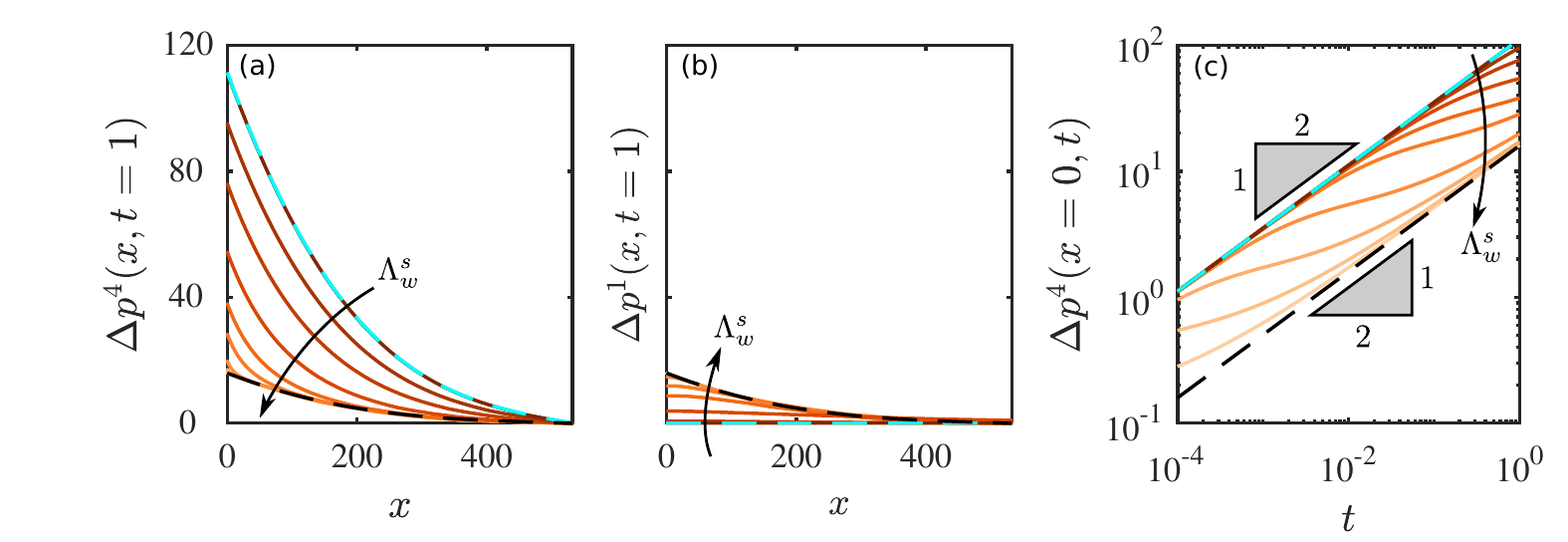}
    \caption{Pressure perturbation during water injection into aquifer~4 of a seven-aquifer system with permeable seals for a wide range of leakage strengths, $\log_{10}(\Lambda_w^s)=-9$, $-7$, $-6.5$, $-6$, $-5.5$, $-5$, $-4$, $-3$, $-2$, and $-1$. We plot the pressure perturbation at $t=1$ against $x$ in (a)~the injection aquifer ($n=4$) and in (b)~the bottom-most aquifer ($n=1$). We also plot (c)~the pressure perturbation at $x=0$ against $t$ on a log scale. All three panels include the analytical solutions for the `no leakage' limit (dashed cyan) and the `strong leakage' limit (dashed black). \label{fig:WI_7L_pressures} }
\end{figure}

We next use our model to investigate the impact of vertical pressure dissipation in more detail. To do so, we solve the water-injection problem described above for a wide range of leakage strengths $\Lambda_w^s$. We find that the pressure in the injection aquifer decreases monotonically with $\Lambda_w^s$, whereas the pressure in all other aquifers increases monotonically with $\Lambda_w^s$ (Figure~\ref{fig:WI_7L_pressures}a--b). Note that compressibility moderates the importance of leakage since increasing compressibility reduces the strength and extent of the pressure perturbation in both $x$ and $z$.

For small values of $\Lambda_w^s$ ($\Lambda_w^s\lesssim{}10^{-7}$), vertical pressure dissipation is unimportant and there is effectively no pressure communication between aquifers. The injection pressure is completely confined to the injection aquifer and evolves according to the solution derived in \S\ref{ss:WI_1L}, such that the pressure perturbation in the injection aquifer evolves as $\Delta{p}^4\sim\sqrt{t}$ (Figure~\ref{fig:WI_7L_pressures}c). The other aquifers are unperturbed by injection, such that $p^n$ remains hydrostatic and $\Delta{p^n}=0$ for $n\neq{}4$. This is the `no-leakage' limit shown in Figure~\ref{fig:WI_7L_pressures}.

For large values of $\Lambda_w^s$ ($\Lambda_w^s\gtrsim{}10^{-3}$), the seals provide essentially no barrier to vertical pressure dissipation. As a result, pressure equilibrates rapidly in the vertical direction and the system behaves like a single aquifer with an effective thickness of $N_zH$, with each aquifer experiencing $1/N_z$ of the total injection rate. The pressures can therefore be described by an effective model,
\begin{equation}\label{eq:lin_diff_strong}
    N_{cw}(R_{cw}+1)\frac{\partial{p^n}}{\partial{t}} -\frac{\partial^2{p^n}}{\partial{x^2}} =\frac{2s_gR_d}{N_z}\,\delta(x)\,u(t)
\end{equation}
for $n=1\ldots{}N_z$, with the same boundary and initial conditions as Equation~\eqref{eq:lin_diff_seven-layer}. This has the analytical solution
\begin{equation}\label{eq:an_sol_ML}
    p^n(x,t) =
    \begin{cases}
        \displaystyle p^0-N_g[n+(n-1)b] +\frac{s_gR_d}{N_z}(x+L_x/2) +\frac{\Omega(x,t)}{N_z} & \text{for} \,\, x\leq0, \\[1em]
        \displaystyle p^0-N_g[n+(n-1)b] -\frac{s_gR_d}{N_z}(x-L_x/2) +\frac{\Omega(x,t)}{N_z} & \text{for} \,\, x \geq 0,
    \end{cases}
\end{equation}
where $\Omega(x,t)$ is given in Equation~\eqref{eq:Omega} above. This is the `strong-leakage' limit shown in Figure~\ref{fig:WI_7L_pressures}. In this limit, the pressure perturbation is the same in all aquifers and evolves according to $\Delta{p}^n\sim{}\sqrt{t}$ (Figure~\ref{fig:WI_7L_pressures}c).

For all nonzero values of $\Lambda_w^s$, the pressure perturbation in the injection aquifer follows the `no leakage' limit at early times before transitioning to the `strong leakage' limit at late times. The latter transition occurs once the pressure perturbation reaches the top and bottom boundaries. Both transitions happen earlier as $\Lambda_w^s$ increases (Figure~\ref{fig:WI_7L_pressures}c). At intermediate times, the pressure is in a transitional state between the two limiting cases. \cite{ChangHesse2013} noted the same departure from the early-time $\sqrt{t}$ scaling due to vertical pressure dissipation, but did not capture the late-time return to a $\sqrt{t}$ scaling because their system was vertically infinite.

\subsection{Gas injection with impermeable seals}\label{ss:GI_1L}

We now consider gas injection into a one-aquifer system ($n=N_z=1$) with impermeable seals ($\Lambda_w^s=0$), in which case our model is equivalent to that of \citet{Mathias2009approx} but for a planar (rather than axisymmetric) geometry and accounting for moderate gas compressibility. To describe gas injection, we take $\dot{M}_g^1=\dot{M}\,u(t)$ and $\dot{M}_w^1=0$. The equation for gas is then Equation~\eqref{eq:GOV_NDg} for $n=1$ and $\mathcal{I}_g^1=2\,\delta(x)\,u(t)$, and the equation for water is Equation~\eqref{eq:GOV_NDw} for $n=1$ and $q_{w,z}^1=q_{w,z}^2=\mathcal{I}_w^1=0$.

\begin{figure}
    \centering
    \includegraphics[width=\textwidth]{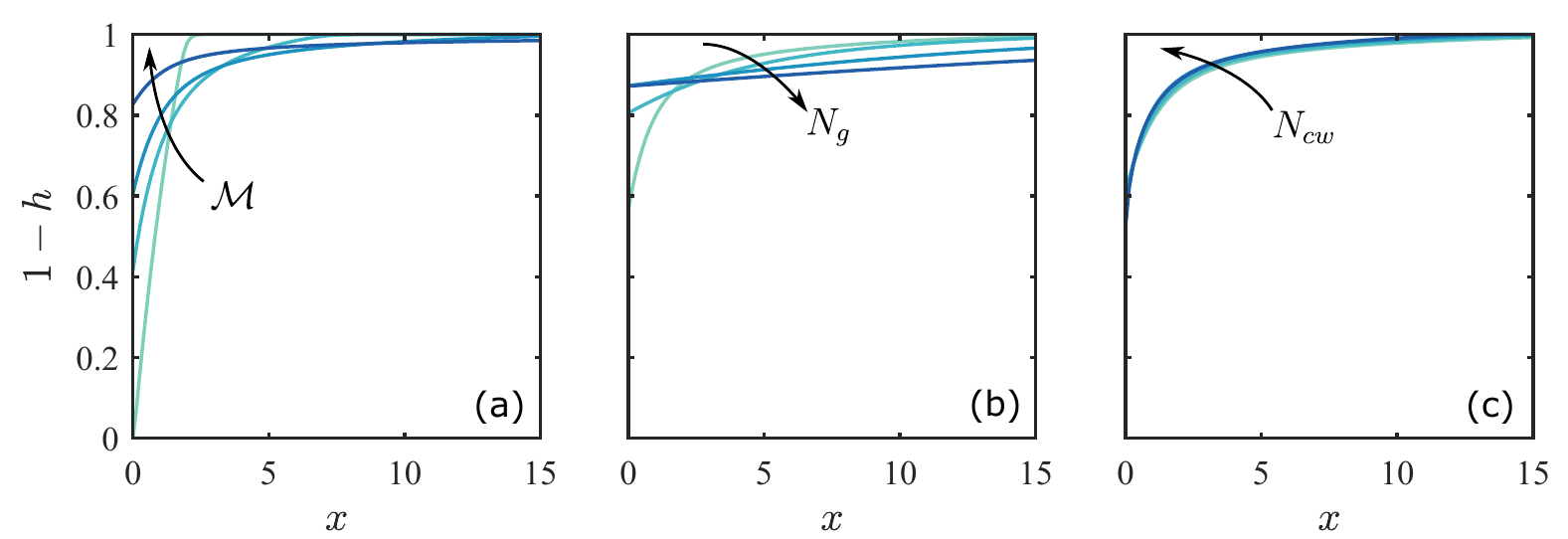}
    \caption{The shape of the gas plume at the end of gas injection into a one-aquifer system with impermeable seals, shown here for  (a)~$\mathcal{M} = 2,\ 10,\ 25\ \& \ 150$, (b)~$N_g = 1,\  20,\ 100\ \& \ 200$, and (c)~$N_{cw} = 3.02 \times 10^{-6},\ 3.02 \times 10^{-5},\ 3.02 \times 10^{-4},\ \& \ 3.02 \times 10^{-3}$. Decreasing $\mathcal{M}$ suppresses tonguing, increasing $N_g$ leads to faster spreading, and increasing $N_{cw}$ makes the gas plume somewhat more compact. The former two parameters have much stronger impacts than the latter. \label{fig:GI_1L_plumes} }
\end{figure}

As it is injected, the gas will spread along the top of the aquifer as a buoyant gravity current. The characteristic tongued shape of the gas plume will be dictated by the interplay between injection pressure, mobility contrast, buoyancy, and compressibility, and is therefore dictated by several different dimensionless parameters: $\mathcal{M}$, $N_g$, $N_{cw}$, $R_{cw}$, $R_{cf}$, $R_d$, and $s_g$. The impacts of these parameters on the shape of the gas plume are, for the most part, well understood from previous work in one-aquifer systems~\citep[\textit{e.g.},][]{Mathias2009approx} and are not the focus of the present study. We illustrate the impacts of $\mathcal{M}$, $N_g$, $N_{cw}$ in Figure~\ref{fig:GI_1L_plumes}. The mobility ratio $\mathcal{M}$ measures the (much higher) mobility of the gas relative to the water and is ultimately responsible for the strongly tongued shape of the gas plume~\citep{nordbotten2006similarity}. Increasing $\mathcal{M}$ increases the severity of this tonguing; decreasing $\mathcal{M}$ suppresses tonguing and focuses the gas near the injection well~(Figure~\ref{fig:GI_1L_plumes}a).
As a pair, density ratio $R_d$ and gravity number $N_g$ measure the importance of buoyancy relative to injection pressure. We expect $R_d<1$ (gas buoyant relative to water), in which case $N_g>1$ implies that the gas will tend to rise and spread significantly due to buoyancy during the injection process, whereas $N_g<1$ implies that buoyancy will play little role during injection~\citep[][and Figure~\ref{fig:GI_1L_plumes}b]{nordbotten2006similarity}. The compressibility number $N_{cw}$ has the weakest impact among these three parameters, despite varying over the largest range. Increasing $N_{cw}$ leads to a slightly more compact and less tongued plume by reducing the strength of the injection pressure and increasing the density of the gas~(Figure~\ref{fig:GI_1L_plumes}c). Although the impact of compressibility on plume shape is relatively small in the scenarios shown here, it can have a stronger effect in other regions of the parameter space~\citep{Mathias2009approx, vilarrasa2010comp}.

\subsection{Gas injection with permeable seals}\label{ss:GI_7L}

Finally, we consider gas injection into the central aquifer ($n=4$) of a seven-aquifer system ($N_z=7$) with permeable seals in order to study the impact of vertical pressure dissipation on the shape of the gas plume. The presence of gas complicates vertical pressure dissipation in the sense that the gas itself presents additional resistance to vertical water flow between the injection aquifer (aquifer~4) and the overlying aquifer (aquifer~5) by obstructing a portion of the seal, and by doing so in the region that is likely to have the highest pressure. Water is likely to be the wetting phase, and may therefore still be able to flow through the gas region via a connected network of residual films. We expect the resistance to this flow to be significantly higher than if the gas were not present. This resistance is quantified by the reduced relative permeability to water in the gas region, $k_{rw}^\star$. Unfortunately, the magnitude of $k_{rw}^\star$ for a network of residual wetting films is very poorly constrained. Although the existence of a connected and conductive network of residual wetting films has been confirmed experimentally~\citep{teige-jgr-2006}, it is not included in standard models for residual saturation and relative permeability. We begin here by considering the interaction of gas injection with pressure dissipation in the case where the gas does not offer any additional resistance to vertical water flow ($k_{rw}^\star=1$). We then consider the more general case of $0\leq{}k_{rw}^\star\leq{}1$, with an emphasis on the most likely scenario of $k_{rw}^\star\ll1$.

\subsubsection{Gas does not obstruct vertical water flow ($k_{rw}^{\star}=1$)}

If the gas provides no additional resistance to vertical flow of water, then we expect gas injection into the central layer of a homogeneous system to lead to a sequence of vertical water fluxes and a pressure distribution that are vertically symmetric across the injection layer---that is, all vertical fluxes should be oriented away from the injection layer and their magnitudes, as well as the pressure in each layer, should depend only on distance from the injection layer. Much like for water injection (\S\ref{ss:WI_7L}), we expect the pressure in the injection aquifer to decrease and the pressure in all other aquifers to increase as $\Lambda_w^s$ increases. We also expect all vertical water fluxes to increase monotonically in magnitude as $\Lambda_w^s$ increases.

\begin{figure}%
    \centering
   \includegraphics[width=\textwidth]{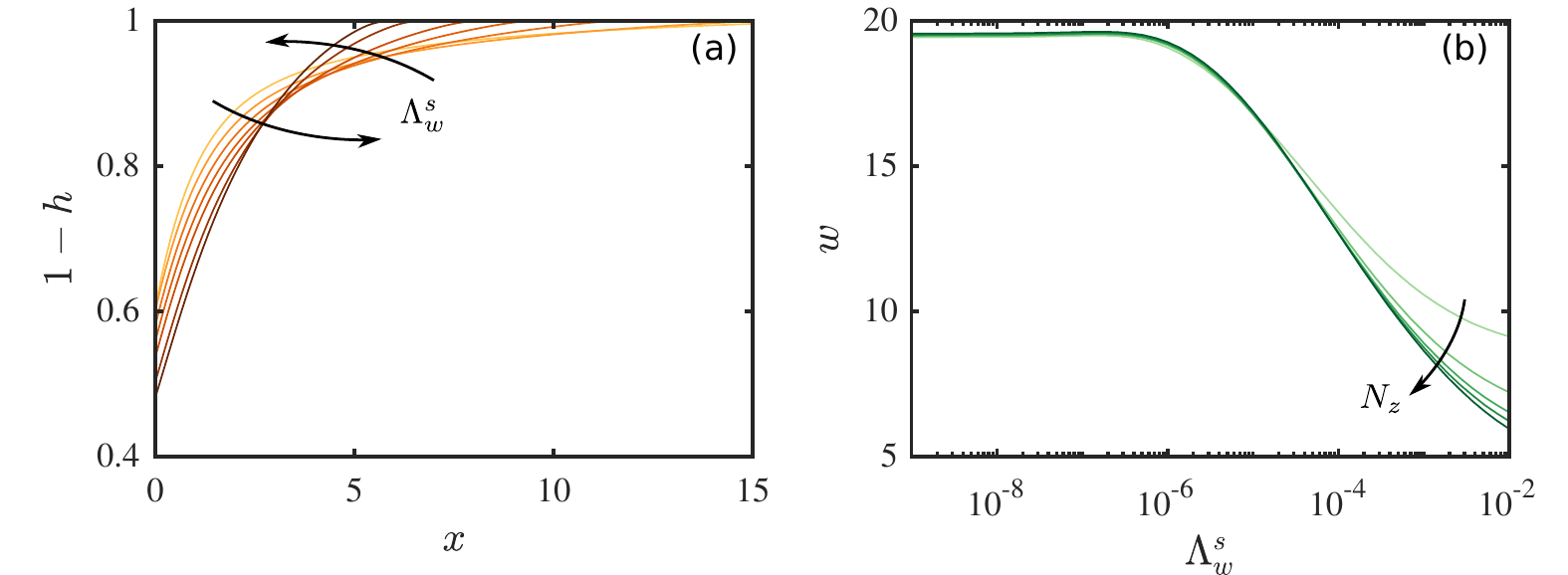}
    \caption{The shape of the gas plume during gas injection into the central aquifer of a seven-aquifer system with permeable seals: (a)~Plume shapes at the end of injection for $\log_{10}\Lambda_w^s = -10$, $-5$, $-4$, $-3.4$, $-3$, $-2.4$, and $-2$, and (b)~plume width $w$ as a function of $\Lambda_w^s$ for injection into systems of $N_z=3$, $5$, $7$, $9$, and $13$ aquifers. \label{fig:GI_7L_plumes} }
\end{figure}

We plot the shape and width of the gas plume for different values of $\Lambda_w^s$ in Figure~\ref{fig:GI_7L_plumes}.
For $\Lambda_w^s\ll{}1$, we reproduce the no-leakage limit from \S\ref{ss:GI_1L}. As $\Lambda_w^s$ increases, we find that the increasingly strong pressure dissipation leads to an increasingly compact plume by suppressing tonguing and thickening the gas column around $x=0$. This is similar to the effect of increasing $N_{cw}$ (Figure~\ref{fig:GI_1L_plumes}), but substantially stronger. To rationalise this behaviour, we consider the impact of vertical pressure dissipation on the pressure gradient driving gas flow.

During injection, the tonguing of the gas plume is driven by the strong pressure gradient and the high mobility of the gas relative to the water. The pressure in the injection layer decreases monotonically with distance from the injection well, and we showed above that it also decreases monotonically with increasing $\Lambda_w^s$; Figure~\ref{fig:WI_7L_pressures}a illustrates these trends for water injection at $t=1$, and gas injection is qualitatively similar. These trends result from the fact that the injected gas must displace water. For $\Lambda_w^s=0$, all of this water is forced laterally through the single injection aquifer, which requires a relatively large pressure gradient. As $\Lambda_w^s$ increases, an increasing fraction of the water is also displaced vertically through the extensive seals and then laterally through other aquifers, thus reducing the pressure gradient in the injection aquifer itself. For strong injection, the lateral gas flow rate is given to a first approximation by $q_{g,x}^n\sim -\lambda_gh^n\partial{p^n}/\partial{x}$ (Equation~\ref{eq:qgx}). As the pressure gradient $\partial{p^n}/\partial{x}$ decreases due to vertical pressure dissipation, the plume thickness $h^n$ must increase in order to achieve the injection rate imposed at $x=0$, thus producing a thicker and more compact gas plume.

To quantify this effect, we measure the width $w$ of the gas plume as a function of $\Lambda_w^s$ and $N_z$, where $w$ is defined as the distance between the injection point and the place where the plume thickness falls below an arbitrary threshold value (here, $10^{-6}$). Note that, in our scaling, a perfectly un-tongued plume (a rectangular block of gas) would have a width of $\sim$$1$. We find that pressure dissipation can decrease the width of the plume by a factor of 2 or more, even for seemingly small values of the leakage strength ($\Lambda_w^s\sim{}10^{-3}$). This effect is amplified by increasing $N_z$, and particularly so for larger values of $\Lambda_w^s$. This effect occurs because pressure dissipation reduces the lateral pressure gradients that drive gas flow (Figure~\ref{fig:WI_7L_pressures}).

\subsubsection{Gas obstructs vertical water flow ($0\leq{}k_{rw}^{\star}\leq{}1$)}

If the gas does provide additional resistance to upward water flow in the injection aquifer, then we expect this resistance to suppress upward pressure dissipation and to enhance downward pressure dissipation, leading to vertical asymmetry in the water fluxes and in the pressure distribution. The importance of this resistance is determined by both $\Lambda_w^s$ and $k_{rw}^{\star}$. In order for the additional resistance from gas in the aquifer to impact pressure dissipation, it must be comparable to (or larger than) the resistance already provided by the seals (roughly, $k_{rw}^{\star}<\Lambda_w^s$).

\begin{figure}%
    \centering
   \includegraphics[width=\textwidth]{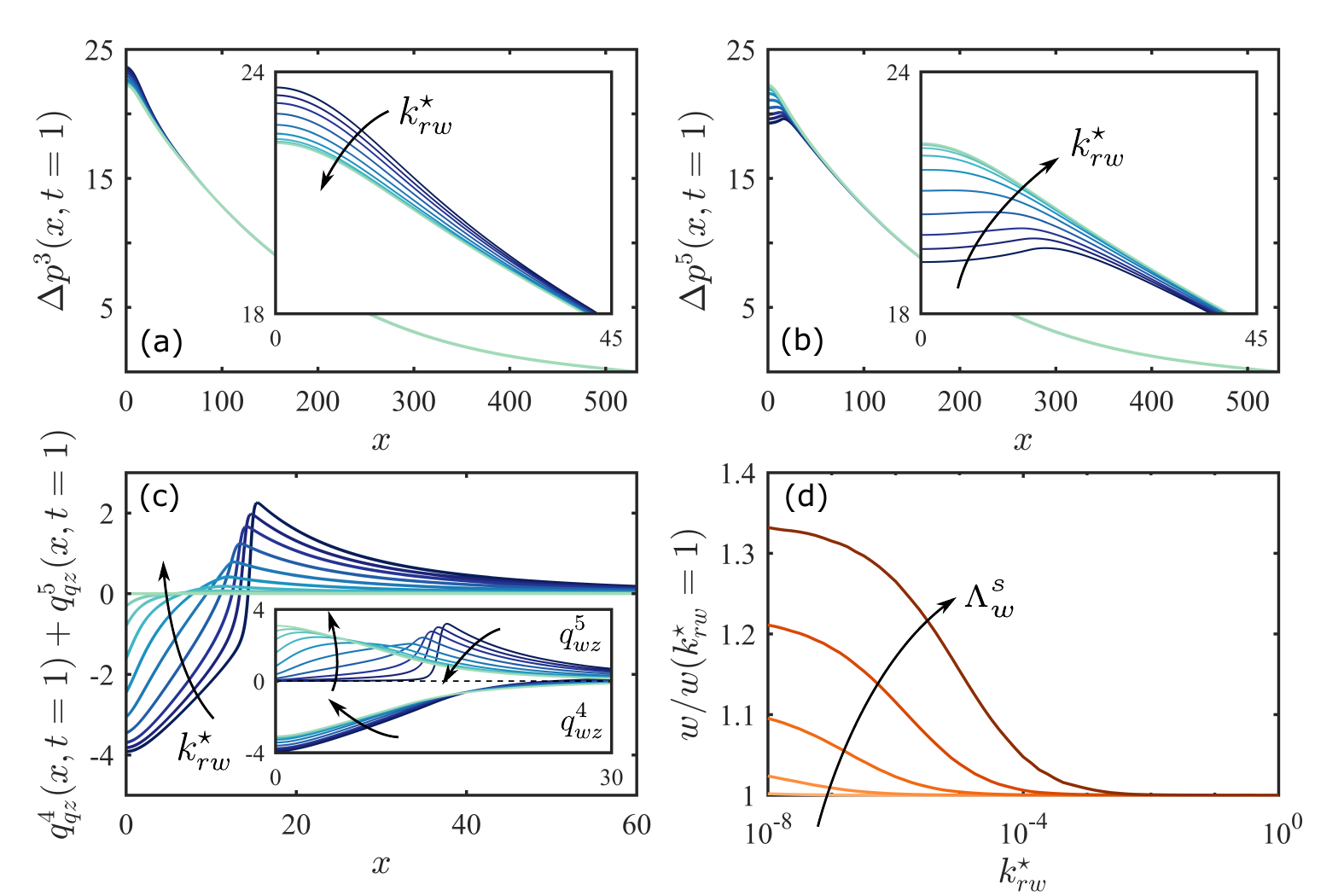}
    \caption{The reduced relative permeability to water within the gas region leads to vertical asymmetry in the pressure field and the vertical water fluxes. Here, we show: The pressure perturbation at $t=1$ in (a)~the aquifer immediately beneath the injection aquifer ($n=3$) and in (b)~the aquifer immediately above the injection aquifer ($n=5$), as well as (c) the net vertical water flux through the injection aquifer ($q_{wz}^4+q_{wz}^5$) at $t=1$ (inset: $q_{wz}^4$ and $q_{wz}^5$ individually). Curves are for $\log_{10}(k_{rw}^{\star})=-8$, $-7$, $-6.5$, $-6$, $-5.5$, $-5$, $-4.5$, $-4$, $-3.5$, and $0$. We also plot (d)~the width of the gas plume at $t=1$ against $k_{rw}^{\star}$, normalised by the plume width for $k_{rw}^{\star}=1$, for $\log_{10}(\Lambda_w^s)=-7$, $-6$, $-5$, $-4$ and $-3$. Note that the horizontal axis of panel (c) is focused near the gas plume. \label{fig:GI_7L_pressures} }
\end{figure}

We illustrate the impact of this resistance in Figure~\ref{fig:GI_7L_pressures}. For injection into aquifer~4, the pressure in the aquifer below (aquifer~3) increases as $k_{rw}^{\star}$ decreases (Fig.~\ref{fig:GI_7L_pressures}a) and the pressure in the aquifer above (aquifer~5) decreases as $k_{rw}^{\star}$ decreases (Fig.~\ref{fig:GI_7L_pressures}b).
For $k_{rw}^\star=1$, the vertical water fluxes through the bottom and top seals of the injection aquifer (seals~4 and 5, respectively) are equal in magnitude and opposite in direction, so they sum to zero (Fig.~\ref{fig:GI_7L_pressures}c, outer plot). As $k_{rw}^\star$ decreases, there is a net downward flow of water immediately under the gas plume and a net upward flow of water elsewhere. The former occurs because the gas obstructs upward flow, as expected; the latter occurs because this obstruction leads to a lower pressure in the aquifer above than in the aquifer below (Fig.~\ref{fig:GI_7L_pressures}a,b), leading to a net upward flow of water in regions unobstructed by gas. Note that all of these effects are localised around the gas plume, and are relatively unimportant in the far field.

Recall that pressure dissipation decreases the width of the gas plume by suppressing tonguing (Fig.~\ref{fig:GI_7L_plumes}). The resistance to flow of water through gas obstructs pressure dissipation and therefore has the opposite effect, increasing the width of the gas plume relative to its width when $k_{rw}^{\star}=1$ (Fig.~\ref{fig:GI_7L_pressures}d).

\section{Discussion \& Conclusions}\label{s:conclusions}

We have developed a new model that couples gas injection and migration with lateral and vertical pressure dissipation in a layered aquifer system. Our model combines a gravity-current representation of the gas with weak vertical flow of water both through the aquifers~\citep{nordbotten2006improved} and across the seals~\citep{hunt-wrr-1985}. Our model constitutes a unique and computationally efficient tool for simultaneously studying the near-field and far-field aspects of gas injection.

Here, we used our model to show that vertical pressure dissipation decreases the pressure in the injection aquifer as well as the width of the gas plume, while increasing the pressure in all other aquifers. For our reference parameters (see table~\ref{tab:params}), the maximum injection pressure and the width of the gas plume are reduced by about two thirds and by about one third, respectively, relative to their values without vertical pressure dissipation ($\Lambda_w^s=0$). Vertical pressure dissipation also slows lateral pressure dissipation, localising pressure buildup around the injection well. These effects have important implications for CCS. The reduction in pressure buildup near the injection well reduces the likelihood of fracturing the caprock. The reduction in lateral pressure dissipation reduces the radius of influence of the injection well, and also the impact of nearby wells and other geological features on injection~\citep{ChangHesse2013}. All of these effects serve to relax the pressure constraint on storage capacity, allowing for longer injection times, larger injection rates, and/or storage in aquifers that are less laterally extensive~\citep{szulczewski-pnas-2012}. In addition, decreasing the width of the CO$_2$ plume by suppressing tonguing leads to a more compact shape at the end of injection, increasing the amount of residual trapping that would occur as the plume later rises, spreads, and migrates~\citep{macminn-compgeosci-2009}. However, a more compact CO$_2$ plume will also have a smaller interfacial area, likely reducing post-injection trapping associated with convective dissolution~\citep[\textit{e.g.},][]{macminn-jfm-2011}.

Although we have developed our model in the context of CO$_2$ injection for CCS, our model can be readily adapted to other subsurface injection problems---for example, waste-water disposal or enhanced oil recovery (EOR). Changing the sign of the source term (\textit{i.e.}, replacing injection with extraction) would allow for exploration of the role of pressure dissipation during hydrocarbon production. We expect vertical pressure dissipation to have a similarly large impact on fluid extraction in a layered system due to its impact on the lateral pressure gradient that draws fluid toward the extraction well.

Vertical pressure dissipation reduces pressure buildup and dissipation in the injection layer~\citep[above and][]{ChangHesse2013}. In addition, the injection pressure does not scale with $\sqrt{t}$ at intermediate times and/or intermediate values of $\Lambda_w^s$. These results have import implications for gas leakage, which occurs when the capillary pressure at the top of the gas plume exceeds the entry pressure of the overlying seal. With negligible vertical water flow, the capillary pressure at the top of the layer is determined by buoyancy and is approximately $p_c \approx (\rho_w-\rho_g)gh$. We have shown that pressure dissipation results in compaction and thickening of the gas plume, which would increase the contribution of buoyancy to capillary pressure. The impacts of vertical water flow and the connectivity of the water through the gas region are less clear. The extension of our model to account for these effects and an exploration of gas leakage will be the subject of future work.


This study was partially funded by the Natural Environment Research Council (NERC) Centre for Doctoral Training (CDT) in Oil \& Gas (grant no. NE/M00578X/1) and funding from Shell International B.V. We are grateful to Lucy Auton for useful discussions and mathematical assistance.


\begin{thebibliography}{36}%
\makeatletter
\providecommand \@ifxundefined [1]{%
 \@ifx{#1\undefined}
}%
\providecommand \@ifnum [1]{%
 \ifnum #1\expandafter \@firstoftwo
 \else \expandafter \@secondoftwo
 \fi
}%
\providecommand \@ifx [1]{%
 \ifx #1\expandafter \@firstoftwo
 \else \expandafter \@secondoftwo
 \fi
}%
\providecommand \natexlab [1]{#1}%
\providecommand \enquote  [1]{``#1''}%
\providecommand \bibnamefont  [1]{#1}%
\providecommand \bibfnamefont [1]{#1}%
\providecommand \citenamefont [1]{#1}%
\providecommand \href@noop [0]{\@secondoftwo}%
\providecommand \href [0]{\begingroup \@sanitize@url \@href}%
\providecommand \@href[1]{\@@startlink{#1}\@@href}%
\providecommand \@@href[1]{\endgroup#1\@@endlink}%
\providecommand \@sanitize@url [0]{\catcode `\\12\catcode `\$12\catcode
  `\&12\catcode `\#12\catcode `\^12\catcode `\_12\catcode `\%12\relax}%
\providecommand \@@startlink[1]{}%
\providecommand \@@endlink[0]{}%
\providecommand \url  [0]{\begingroup\@sanitize@url \@url }%
\providecommand \@url [1]{\endgroup\@href {#1}{\urlprefix }}%
\providecommand \urlprefix  [0]{URL }%
\providecommand \Eprint [0]{\href }%
\providecommand \doibase [0]{http://dx.doi.org/}%
\providecommand \selectlanguage [0]{\@gobble}%
\providecommand \bibinfo  [0]{\@secondoftwo}%
\providecommand \bibfield  [0]{\@secondoftwo}%
\providecommand \translation [1]{[#1]}%
\providecommand \BibitemOpen [0]{}%
\providecommand \bibitemStop [0]{}%
\providecommand \bibitemNoStop [0]{.\EOS\space}%
\providecommand \EOS [0]{\spacefactor3000\relax}%
\providecommand \BibitemShut  [1]{\csname bibitem#1\endcsname}%
\let\auto@bib@innerbib\@empty
\bibitem [{\citenamefont {IPCC}(2005)}]{ipcc-cambridge-2005}%
  \BibitemOpen
  \bibfield  {author} {\bibinfo {author} {\bibnamefont {IPCC}},\ }\href@noop {}
  {\emph {\bibinfo {title} {Carbon {D}ioxide {C}apture and {S}torage}}},\
  \bibinfo {type} {Special {R}eport Prepared by {W}orking {G}roup {III} of the
  {I}ntergovernmental {P}anel on {C}limate {C}hange}\ (\bibinfo {address}
  {Cambridge, UK},\ \bibinfo {year} {2005})\BibitemShut {NoStop}%
\bibitem [{\citenamefont {Szulczewski}\ \emph {et~al.}(2012)\citenamefont
  {Szulczewski}, \citenamefont {MacMinn}, \citenamefont {Herzog},\ and\
  \citenamefont {Juanes}}]{szulczewski-pnas-2012}%
  \BibitemOpen
  \bibfield  {author} {\bibinfo {author} {\bibfnamefont {M.~L.}\ \bibnamefont
  {Szulczewski}}, \bibinfo {author} {\bibfnamefont {C.~W.}\ \bibnamefont
  {MacMinn}}, \bibinfo {author} {\bibfnamefont {H.~J.}\ \bibnamefont {Herzog}},
  \ and\ \bibinfo {author} {\bibfnamefont {R.}~\bibnamefont {Juanes}},\
  }\bibfield  {title} {\enquote {\bibinfo {title} {Lifetime of carbon capture
  and storage as a climate-change mitigation technology},}\ }\href@noop {}
  {\bibfield  {journal} {\bibinfo  {journal} {Proceedings of the National
  Academy of Sciences of the United States of America}\ }\textbf {\bibinfo
  {volume} {109}},\ \bibinfo {pages} {5185--5189} (\bibinfo {year}
  {2012})}\BibitemShut {NoStop}%
\bibitem [{\citenamefont {Huppert}\ and\ \citenamefont
  {Woods}(1995)}]{huppert-jfm-1995}%
  \BibitemOpen
  \bibfield  {author} {\bibinfo {author} {\bibfnamefont {H.~E.}\ \bibnamefont
  {Huppert}}\ and\ \bibinfo {author} {\bibfnamefont {A.~W.}\ \bibnamefont
  {Woods}},\ }\bibfield  {title} {\enquote {\bibinfo {title} {Gravity-driven
  flows in porous layers},}\ }\href@noop {} {\bibfield  {journal} {\bibinfo
  {journal} {Journal of Fluid Mechanics}\ }\textbf {\bibinfo {volume} {292}},\
  \bibinfo {pages} {55--69} (\bibinfo {year} {1995})}\BibitemShut {NoStop}%
\bibitem [{\citenamefont {Nordbotten}\ \emph {et~al.}(2005)\citenamefont
  {Nordbotten}, \citenamefont {Celia},\ and\ \citenamefont
  {Bachu}}]{nordbotten2005injection}%
  \BibitemOpen
  \bibfield  {author} {\bibinfo {author} {\bibfnamefont {Jan~Martin}\
  \bibnamefont {Nordbotten}}, \bibinfo {author} {\bibfnamefont {Michael~A}\
  \bibnamefont {Celia}}, \ and\ \bibinfo {author} {\bibfnamefont {Stefan}\
  \bibnamefont {Bachu}},\ }\bibfield  {title} {\enquote {\bibinfo {title}
  {Injection and storage of {CO}$_2$ in deep saline aquifers: Analytical
  solution for {CO}$_2$ plume evolution during injection},}\ }\href@noop {}
  {\bibfield  {journal} {\bibinfo  {journal} {Transport in Porous media}\
  }\textbf {\bibinfo {volume} {58}},\ \bibinfo {pages} {339--360} (\bibinfo
  {year} {2005})}\BibitemShut {NoStop}%
\bibitem [{\citenamefont {Nordbotten}\ and\ \citenamefont
  {Celia}(2006{\natexlab{a}})}]{nordbotten2006similarity}%
  \BibitemOpen
  \bibfield  {author} {\bibinfo {author} {\bibfnamefont {Jan~M}\ \bibnamefont
  {Nordbotten}}\ and\ \bibinfo {author} {\bibfnamefont {Michael~A}\
  \bibnamefont {Celia}},\ }\bibfield  {title} {\enquote {\bibinfo {title}
  {Similarity solutions for fluid injection into confined aquifers},}\
  }\href@noop {} {\bibfield  {journal} {\bibinfo  {journal} {Journal of Fluid
  Mechanics}\ }\textbf {\bibinfo {volume} {561}},\ \bibinfo {pages} {307--327}
  (\bibinfo {year} {2006}{\natexlab{a}})}\BibitemShut {NoStop}%
\bibitem [{\citenamefont {Hesse}\ \emph {et~al.}(2007)\citenamefont {Hesse},
  \citenamefont {Tchelepi}, \citenamefont {Cantwell},\ and\ \citenamefont
  {Orr~Jr.}}]{hesse-jfm-2007}%
  \BibitemOpen
  \bibfield  {author} {\bibinfo {author} {\bibfnamefont {M.~A.}\ \bibnamefont
  {Hesse}}, \bibinfo {author} {\bibfnamefont {H.~A.}\ \bibnamefont {Tchelepi}},
  \bibinfo {author} {\bibfnamefont {B.~J.}\ \bibnamefont {Cantwell}}, \ and\
  \bibinfo {author} {\bibfnamefont {F.~M.}\ \bibnamefont {Orr~Jr.}},\
  }\bibfield  {title} {\enquote {\bibinfo {title} {Gravity currents in
  horizontal porous layers: transition from early to late self-similarity},}\
  }\href@noop {} {\bibfield  {journal} {\bibinfo  {journal} {Journal of Fluid
  Mechanics}\ }\textbf {\bibinfo {volume} {577}},\ \bibinfo {pages} {363--383}
  (\bibinfo {year} {2007})}\BibitemShut {NoStop}%
\bibitem [{\citenamefont {Gasda}\ \emph {et~al.}(2009)\citenamefont {Gasda},
  \citenamefont {Nordbotten},\ and\ \citenamefont
  {Celia}}]{gasda-compgeosci-2009}%
  \BibitemOpen
  \bibfield  {author} {\bibinfo {author} {\bibfnamefont {S.~E.}\ \bibnamefont
  {Gasda}}, \bibinfo {author} {\bibfnamefont {J.~M.}\ \bibnamefont
  {Nordbotten}}, \ and\ \bibinfo {author} {\bibfnamefont {M.~A.}\ \bibnamefont
  {Celia}},\ }\bibfield  {title} {\enquote {\bibinfo {title} {Vertical
  equilibrium with sub-scale analytical methods for geological {CO}$_2$
  sequestration},}\ }\href@noop {} {\bibfield  {journal} {\bibinfo  {journal}
  {Computational Geosciences}\ }\textbf {\bibinfo {volume} {79}},\ \bibinfo
  {pages} {15--27} (\bibinfo {year} {2009})}\BibitemShut {NoStop}%
\bibitem [{\citenamefont {Juanes}\ \emph {et~al.}(2010)\citenamefont {Juanes},
  \citenamefont {MacMinn},\ and\ \citenamefont
  {Szulczewski}}]{juanes-tpm-2010}%
  \BibitemOpen
  \bibfield  {author} {\bibinfo {author} {\bibfnamefont {R.}~\bibnamefont
  {Juanes}}, \bibinfo {author} {\bibfnamefont {C.~W.}\ \bibnamefont {MacMinn}},
  \ and\ \bibinfo {author} {\bibfnamefont {M.~L.}\ \bibnamefont
  {Szulczewski}},\ }\bibfield  {title} {\enquote {\bibinfo {title} {The
  footprint of the {CO}$_2$ plume during carbon dioxide storage in saline
  aquifers: {S}torage efficiency for capillary trapping at the basin scale},}\
  }\href@noop {} {\bibfield  {journal} {\bibinfo  {journal} {Transport in
  Porous Media}\ }\textbf {\bibinfo {volume} {82}},\ \bibinfo {pages} {19--30}
  (\bibinfo {year} {2010})}\BibitemShut {NoStop}%
\bibitem [{\citenamefont {Mathias}\ \emph {et~al.}(2009)\citenamefont
  {Mathias}, \citenamefont {Hardisty}, \citenamefont {Trudell},\ and\
  \citenamefont {Zimmerman}}]{Mathias2009approx}%
  \BibitemOpen
  \bibfield  {author} {\bibinfo {author} {\bibfnamefont {Simon~A}\ \bibnamefont
  {Mathias}}, \bibinfo {author} {\bibfnamefont {Paul~E}\ \bibnamefont
  {Hardisty}}, \bibinfo {author} {\bibfnamefont {Mark~R}\ \bibnamefont
  {Trudell}}, \ and\ \bibinfo {author} {\bibfnamefont {Robert~W}\ \bibnamefont
  {Zimmerman}},\ }\bibfield  {title} {\enquote {\bibinfo {title} {Approximate
  solutions for pressure buildup during {CO}$_2$ injection in brine
  aquifers},}\ }\href@noop {} {\bibfield  {journal} {\bibinfo  {journal}
  {Transport in Porous Media}\ }\textbf {\bibinfo {volume} {79}},\ \bibinfo
  {pages} {265--284} (\bibinfo {year} {2009})}\BibitemShut {NoStop}%
\bibitem [{\citenamefont {MacMinn}\ \emph {et~al.}(2010)\citenamefont
  {MacMinn}, \citenamefont {Szulczewski},\ and\ \citenamefont
  {Juanes}}]{macminn2010co2}%
  \BibitemOpen
  \bibfield  {author} {\bibinfo {author} {\bibfnamefont {Christopher~W}\
  \bibnamefont {MacMinn}}, \bibinfo {author} {\bibfnamefont {Michael~Lawrence}\
  \bibnamefont {Szulczewski}}, \ and\ \bibinfo {author} {\bibfnamefont {Ruben}\
  \bibnamefont {Juanes}},\ }\bibfield  {title} {\enquote {\bibinfo {title}
  {{CO}$_2$ migration in saline aquifers. part 1. capillary trapping under
  slope and groundwater flow},}\ }\href@noop {} {\bibfield  {journal} {\bibinfo
   {journal} {Journal of fluid mechanics}\ }\textbf {\bibinfo {volume} {662}},\
  \bibinfo {pages} {329--351} (\bibinfo {year} {2010})}\BibitemShut {NoStop}%
\bibitem [{\citenamefont {Pegler}\ \emph {et~al.}(2014)\citenamefont {Pegler},
  \citenamefont {Huppert},\ and\ \citenamefont
  {Neufeld}}]{Pegler2014injection}%
  \BibitemOpen
  \bibfield  {author} {\bibinfo {author} {\bibfnamefont {Samuel~S}\
  \bibnamefont {Pegler}}, \bibinfo {author} {\bibfnamefont {Herbert~E}\
  \bibnamefont {Huppert}}, \ and\ \bibinfo {author} {\bibfnamefont {Jerome~A}\
  \bibnamefont {Neufeld}},\ }\bibfield  {title} {\enquote {\bibinfo {title}
  {Fluid injection into a confined porous layer},}\ }\href@noop {} {\bibfield
  {journal} {\bibinfo  {journal} {Journal of Fluid Mechanics}\ }\textbf
  {\bibinfo {volume} {745}},\ \bibinfo {pages} {592--620} (\bibinfo {year}
  {2014})}\BibitemShut {NoStop}%
\bibitem [{\citenamefont {Zheng}\ \emph {et~al.}(2015)\citenamefont {Zheng},
  \citenamefont {Guo}, \citenamefont {Christov}, \citenamefont {Celia},\ and\
  \citenamefont {Stone}}]{zheng-jfm-2015}%
  \BibitemOpen
  \bibfield  {author} {\bibinfo {author} {\bibfnamefont {Z.}~\bibnamefont
  {Zheng}}, \bibinfo {author} {\bibfnamefont {B.}~\bibnamefont {Guo}}, \bibinfo
  {author} {\bibfnamefont {I.~C.}\ \bibnamefont {Christov}}, \bibinfo {author}
  {\bibfnamefont {M.~A.}\ \bibnamefont {Celia}}, \ and\ \bibinfo {author}
  {\bibfnamefont {H.~A.}\ \bibnamefont {Stone}},\ }\bibfield  {title} {\enquote
  {\bibinfo {title} {Flow regimes for fluid injection into a confined porous
  medium},}\ }\href@noop {} {\bibfield  {journal} {\bibinfo  {journal} {Journal
  of Fluid Mechanics}\ }\textbf {\bibinfo {volume} {767}},\ \bibinfo {pages}
  {881--909} (\bibinfo {year} {2015})}\BibitemShut {NoStop}%
\bibitem [{\citenamefont {Golding}\ \emph {et~al.}(2017)\citenamefont
  {Golding}, \citenamefont {Huppert},\ and\ \citenamefont
  {Neufeld}}]{golding-jfm-2017}%
  \BibitemOpen
  \bibfield  {author} {\bibinfo {author} {\bibfnamefont {M.~J.}\ \bibnamefont
  {Golding}}, \bibinfo {author} {\bibfnamefont {H.~E.}\ \bibnamefont
  {Huppert}}, \ and\ \bibinfo {author} {\bibfnamefont {J.~A.}\ \bibnamefont
  {Neufeld}},\ }\bibfield  {title} {\enquote {\bibinfo {title} {Two-phase
  gravity currents resulting from the release of a fixed volume of fluid in a
  porous medium},}\ }\href@noop {} {\bibfield  {journal} {\bibinfo  {journal}
  {Journal of Fluid Mechanics}\ }\textbf {\bibinfo {volume} {832}},\ \bibinfo
  {pages} {550---577} (\bibinfo {year} {2017})}\BibitemShut {NoStop}%
\bibitem [{\citenamefont {Mathias}\ \emph {et~al.}(2011)\citenamefont
  {Mathias}, \citenamefont {{Gonz\'{a}lez Mart\'{i}nez de Miguel}},
  \citenamefont {Thatcher},\ and\ \citenamefont
  {Zimmerman}}]{mathias-tipm-2011}%
  \BibitemOpen
  \bibfield  {author} {\bibinfo {author} {\bibfnamefont {S.~A.}\ \bibnamefont
  {Mathias}}, \bibinfo {author} {\bibfnamefont {G.~J.}\ \bibnamefont
  {{Gonz\'{a}lez Mart\'{i}nez de Miguel}}}, \bibinfo {author} {\bibfnamefont
  {K.~E.}\ \bibnamefont {Thatcher}}, \ and\ \bibinfo {author} {\bibfnamefont
  {R.~W.}\ \bibnamefont {Zimmerman}},\ }\bibfield  {title} {\enquote {\bibinfo
  {title} {Pressure buildup during {CO}$_2$ injection into a closed brine
  aquifer},}\ }\href@noop {} {\bibfield  {journal} {\bibinfo  {journal}
  {Transport in Porous Media}\ }\textbf {\bibinfo {volume} {89}},\ \bibinfo
  {pages} {383--397} (\bibinfo {year} {2011})}\BibitemShut {NoStop}%
\bibitem [{\citenamefont {Hewitt}\ \emph {et~al.}(2015)\citenamefont {Hewitt},
  \citenamefont {Neufeld},\ and\ \citenamefont {Balmforth}}]{hewitt-jfm-2015}%
  \BibitemOpen
  \bibfield  {author} {\bibinfo {author} {\bibfnamefont {D.~R.}\ \bibnamefont
  {Hewitt}}, \bibinfo {author} {\bibfnamefont {J.~A.}\ \bibnamefont {Neufeld}},
  \ and\ \bibinfo {author} {\bibfnamefont {N.~J.}\ \bibnamefont {Balmforth}},\
  }\bibfield  {title} {\enquote {\bibinfo {title} {Shallow, gravity-driven flow
  in a poro-elastic layer},}\ }\href@noop {} {\bibfield  {journal} {\bibinfo
  {journal} {Journal of Fluid Mechanics}\ }\textbf {\bibinfo {volume} {778}},\
  \bibinfo {pages} {335--360} (\bibinfo {year} {2015})}\BibitemShut {NoStop}%
\bibitem [{\citenamefont {Nicot}(2008)}]{nicot-ijggc-2008}%
  \BibitemOpen
  \bibfield  {author} {\bibinfo {author} {\bibfnamefont {J.-P.}\ \bibnamefont
  {Nicot}},\ }\bibfield  {title} {\enquote {\bibinfo {title} {Evaluation of
  large-scale {CO}$_2$ storage on fresh-water sections of aquifers: An example
  from the {T}exas {G}ulf {C}oast {B}asin},}\ }\href@noop {} {\bibfield
  {journal} {\bibinfo  {journal} {International Journal of Greenhouse Gas
  Control}\ }\textbf {\bibinfo {volume} {2}},\ \bibinfo {pages} {582--593}
  (\bibinfo {year} {2008})}\BibitemShut {NoStop}%
\bibitem [{\citenamefont {Birkholzer}\ \emph {et~al.}(2009)\citenamefont
  {Birkholzer}, \citenamefont {Zhou},\ and\ \citenamefont
  {Tsang}}]{birkholzer2009large}%
  \BibitemOpen
  \bibfield  {author} {\bibinfo {author} {\bibfnamefont {Jens~T}\ \bibnamefont
  {Birkholzer}}, \bibinfo {author} {\bibfnamefont {Quanlin}\ \bibnamefont
  {Zhou}}, \ and\ \bibinfo {author} {\bibfnamefont {Chin-Fu}\ \bibnamefont
  {Tsang}},\ }\bibfield  {title} {\enquote {\bibinfo {title} {Large-scale
  impact of {CO}$_2$ storage in deep saline aquifers: A sensitivity study on
  pressure response in stratified systems},}\ }\href@noop {} {\bibfield
  {journal} {\bibinfo  {journal} {International Journal of Greenhouse Gas
  Control}\ }\textbf {\bibinfo {volume} {3}},\ \bibinfo {pages} {181--194}
  (\bibinfo {year} {2009})}\BibitemShut {NoStop}%
\bibitem [{\citenamefont {Chang}\ \emph {et~al.}(2013)\citenamefont {Chang},
  \citenamefont {Hesse},\ and\ \citenamefont {Nicot}}]{ChangHesse2013}%
  \BibitemOpen
  \bibfield  {author} {\bibinfo {author} {\bibfnamefont {Kyung~Won}\
  \bibnamefont {Chang}}, \bibinfo {author} {\bibfnamefont {Marc~A.}\
  \bibnamefont {Hesse}}, \ and\ \bibinfo {author} {\bibfnamefont
  {Jean‐Philippe}\ \bibnamefont {Nicot}},\ }\bibfield  {title} {\enquote
  {\bibinfo {title} {Reduction of lateral pressure propagation due to
  dissipation into ambient mudrocks during geological carbon dioxide
  storage},}\ }\href@noop {} {\bibfield  {journal} {\bibinfo  {journal} {Water
  Resources Research}\ }\textbf {\bibinfo {volume} {49}},\ \bibinfo {pages}
  {2573--2588} (\bibinfo {year} {2013})}\BibitemShut {NoStop}%
\bibitem [{\citenamefont {Nicot}\ \emph {et~al.}(2011)\citenamefont {Nicot},
  \citenamefont {Hosseini},\ and\ \citenamefont {Solano}}]{nicot-ghgt-2010}%
  \BibitemOpen
  \bibfield  {author} {\bibinfo {author} {\bibfnamefont {J.-P.}\ \bibnamefont
  {Nicot}}, \bibinfo {author} {\bibfnamefont {S.~A.}\ \bibnamefont {Hosseini}},
  \ and\ \bibinfo {author} {\bibfnamefont {S.~V.}\ \bibnamefont {Solano}},\
  }\bibfield  {title} {\enquote {\bibinfo {title} {Are single-phase flow
  numerical models sufficient to estimate pressure distribution in {CO}$_2$
  sequestration projects?}}\ }\href@noop {} {\bibfield  {journal} {\bibinfo
  {journal} {Energy Procedia (Proceedings of the 10th International Conference
  on Greenhouse Gas Control Technologies)}\ }\textbf {\bibinfo {volume} {4}},\
  \bibinfo {pages} {3919--3926} (\bibinfo {year} {2011})}\BibitemShut {NoStop}%
\bibitem [{\citenamefont {Golding}\ \emph {et~al.}(2011)\citenamefont
  {Golding}, \citenamefont {Neufeld}, \citenamefont {Hesse},\ and\
  \citenamefont {Huppert}}]{Golding2011}%
  \BibitemOpen
  \bibfield  {author} {\bibinfo {author} {\bibfnamefont {Madeleine~J}\
  \bibnamefont {Golding}}, \bibinfo {author} {\bibfnamefont {Jerome~A}\
  \bibnamefont {Neufeld}}, \bibinfo {author} {\bibfnamefont {Marc~A}\
  \bibnamefont {Hesse}}, \ and\ \bibinfo {author} {\bibfnamefont {Herbert~E}\
  \bibnamefont {Huppert}},\ }\bibfield  {title} {\enquote {\bibinfo {title}
  {Two-phase gravity currents in porous media},}\ }\href@noop {} {\bibfield
  {journal} {\bibinfo  {journal} {Journal of Fluid Mechanics}\ }\textbf
  {\bibinfo {volume} {678}},\ \bibinfo {pages} {248--270} (\bibinfo {year}
  {2011})}\BibitemShut {NoStop}%
\bibitem [{\citenamefont {Kochina}\ \emph {et~al.}(1983)\citenamefont
  {Kochina}, \citenamefont {Mikhailov},\ and\ \citenamefont
  {Filinov}}]{kochina-intjengsci-1983}%
  \BibitemOpen
  \bibfield  {author} {\bibinfo {author} {\bibfnamefont {I.~N.}\ \bibnamefont
  {Kochina}}, \bibinfo {author} {\bibfnamefont {N.~N.}\ \bibnamefont
  {Mikhailov}}, \ and\ \bibinfo {author} {\bibfnamefont {M.~V.}\ \bibnamefont
  {Filinov}},\ }\bibfield  {title} {\enquote {\bibinfo {title} {Groundwater
  mound damping},}\ }\href@noop {} {\bibfield  {journal} {\bibinfo  {journal}
  {International Journal of Engineering Science}\ }\textbf {\bibinfo {volume}
  {21}},\ \bibinfo {pages} {413--421} (\bibinfo {year} {1983})}\BibitemShut
  {NoStop}%
\bibitem [{\citenamefont {Barenblatt}(1996)}]{barenblatt-cambridge-1996}%
  \BibitemOpen
  \bibfield  {author} {\bibinfo {author} {\bibfnamefont {G.~I.}\ \bibnamefont
  {Barenblatt}},\ }\href@noop {} {\emph {\bibinfo {title} {Scaling,
  self-similarity, and intermediate asymptotics}}}\ (\bibinfo  {publisher}
  {Cambridge University Press},\ \bibinfo {year} {1996})\BibitemShut {NoStop}%
\bibitem [{\citenamefont {Hesse}\ \emph {et~al.}(2008)\citenamefont {Hesse},
  \citenamefont {Orr~Jr.},\ and\ \citenamefont {Tchelepi}}]{hesse-jfm-2008}%
  \BibitemOpen
  \bibfield  {author} {\bibinfo {author} {\bibfnamefont {M.~A.}\ \bibnamefont
  {Hesse}}, \bibinfo {author} {\bibfnamefont {F.~M.}\ \bibnamefont {Orr~Jr.}},
  \ and\ \bibinfo {author} {\bibfnamefont {H.~A.}\ \bibnamefont {Tchelepi}},\
  }\bibfield  {title} {\enquote {\bibinfo {title} {Gravity currents with
  residual trapping},}\ }\href@noop {} {\bibfield  {journal} {\bibinfo
  {journal} {Journal of Fluid Mechanics}\ }\textbf {\bibinfo {volume} {611}},\
  \bibinfo {pages} {35--60} (\bibinfo {year} {2008})}\BibitemShut {NoStop}%
\bibitem [{\citenamefont {Yortsos}(1995)}]{yortsos-tpm-1995}%
  \BibitemOpen
  \bibfield  {author} {\bibinfo {author} {\bibfnamefont {Y.~C.}\ \bibnamefont
  {Yortsos}},\ }\bibfield  {title} {\enquote {\bibinfo {title} {A theoretical
  analysis of vertical flow equilibrium},}\ }\href@noop {} {\bibfield
  {journal} {\bibinfo  {journal} {Transport in Porous Media}\ }\textbf
  {\bibinfo {volume} {18}},\ \bibinfo {pages} {107--129} (\bibinfo {year}
  {1995})}\BibitemShut {NoStop}%
\bibitem [{\citenamefont {de~Loubens}\ and\ \citenamefont
  {Ramakrishnan}(2011)}]{deloubens-jfm-2011}%
  \BibitemOpen
  \bibfield  {author} {\bibinfo {author} {\bibfnamefont {R.}~\bibnamefont
  {de~Loubens}}\ and\ \bibinfo {author} {\bibfnamefont {T.~S.}\ \bibnamefont
  {Ramakrishnan}},\ }\bibfield  {title} {\enquote {\bibinfo {title} {Analysis
  and computation of gravity-induced migration in porous media},}\ }\href@noop
  {} {\bibfield  {journal} {\bibinfo  {journal} {Journal of Fluid Mechanics}\
  }\textbf {\bibinfo {volume} {675}},\ \bibinfo {pages} {60--86} (\bibinfo
  {year} {2011})}\BibitemShut {NoStop}%
\bibitem [{\citenamefont {Nordbotten}\ and\ \citenamefont
  {Celia}(2006{\natexlab{b}})}]{nordbotten2006improved}%
  \BibitemOpen
  \bibfield  {author} {\bibinfo {author} {\bibfnamefont {Jan~M}\ \bibnamefont
  {Nordbotten}}\ and\ \bibinfo {author} {\bibfnamefont {Michael~A}\
  \bibnamefont {Celia}},\ }\bibfield  {title} {\enquote {\bibinfo {title} {An
  improved analytical solution for interface upconing around a well},}\
  }\href@noop {} {\bibfield  {journal} {\bibinfo  {journal} {Water Resources
  Research}\ }\textbf {\bibinfo {volume} {42}} (\bibinfo {year}
  {2006}{\natexlab{b}})}\BibitemShut {NoStop}%
\bibitem [{\citenamefont {Bear}(1972)}]{Bear1972}%
  \BibitemOpen
  \bibfield  {author} {\bibinfo {author} {\bibfnamefont {Jacob}\ \bibnamefont
  {Bear}},\ }\href@noop {} {\emph {\bibinfo {title} {Dynamics of fluids in
  porous media}}}\ (\bibinfo  {publisher} {Courier Corporation},\ \bibinfo
  {year} {1972})\BibitemShut {NoStop}%
\bibitem [{\citenamefont {{van der Kamp}}\ and\ \citenamefont
  {Gale}(1983)}]{vanderkamp-wrr-1983}%
  \BibitemOpen
  \bibfield  {author} {\bibinfo {author} {\bibfnamefont {G.}~\bibnamefont {{van
  der Kamp}}}\ and\ \bibinfo {author} {\bibfnamefont {J.~E.}\ \bibnamefont
  {Gale}},\ }\bibfield  {title} {\enquote {\bibinfo {title} {Theory of {Earth}
  tide and barometric effects in porous formations with compressible grains},}\
  }\href@noop {} {\bibfield  {journal} {\bibinfo  {journal} {Water Resources
  Research}\ }\textbf {\bibinfo {volume} {19}},\ \bibinfo {pages} {538--544}
  (\bibinfo {year} {1983})}\BibitemShut {NoStop}%
\bibitem [{\citenamefont {Green}\ and\ \citenamefont
  {Wang}(1990)}]{green-wrr-1990}%
  \BibitemOpen
  \bibfield  {author} {\bibinfo {author} {\bibfnamefont {D.~H.}\ \bibnamefont
  {Green}}\ and\ \bibinfo {author} {\bibfnamefont {H.~F.}\ \bibnamefont
  {Wang}},\ }\bibfield  {title} {\enquote {\bibinfo {title} {Specific storage
  as a poroelastic coefficient},}\ }\href@noop {} {\bibfield  {journal}
  {\bibinfo  {journal} {Water Resources Research}\ }\textbf {\bibinfo {volume}
  {26}},\ \bibinfo {pages} {1631--1637} (\bibinfo {year} {1990})}\BibitemShut
  {NoStop}%
\bibitem [{\citenamefont {Bear}(1979)}]{Bear1979}%
  \BibitemOpen
  \bibfield  {author} {\bibinfo {author} {\bibfnamefont {Jacob}\ \bibnamefont
  {Bear}},\ }\href@noop {} {\emph {\bibinfo {title} {Hydraulics of
  groundwater}}}\ (\bibinfo  {publisher} {Courier Corporation},\ \bibinfo
  {year} {1979})\BibitemShut {NoStop}%
\bibitem [{\citenamefont {Hunt}(1985)}]{hunt-wrr-1985}%
  \BibitemOpen
  \bibfield  {author} {\bibinfo {author} {\bibfnamefont {B.}~\bibnamefont
  {Hunt}},\ }\bibfield  {title} {\enquote {\bibinfo {title} {Flow to a well in
  a multiaquifer system},}\ }\href@noop {} {\bibfield  {journal} {\bibinfo
  {journal} {Water Resources Research}\ }\textbf {\bibinfo {volume} {21}},\
  \bibinfo {pages} {1637--1641} (\bibinfo {year} {1985})}\BibitemShut {NoStop}%
\bibitem [{\citenamefont {Shampine}\ and\ \citenamefont
  {Reichelt}(1997)}]{matlab}%
  \BibitemOpen
  \bibfield  {author} {\bibinfo {author} {\bibfnamefont {Lawrence~F}\
  \bibnamefont {Shampine}}\ and\ \bibinfo {author} {\bibfnamefont {Mark~W}\
  \bibnamefont {Reichelt}},\ }\bibfield  {title} {\enquote {\bibinfo {title}
  {The matlab ode suite},}\ }\href@noop {} {\bibfield  {journal} {\bibinfo
  {journal} {SIAM journal on scientific computing}\ }\textbf {\bibinfo {volume}
  {18}},\ \bibinfo {pages} {1--22} (\bibinfo {year} {1997})}\BibitemShut
  {NoStop}%
\bibitem [{\citenamefont {Vilarrasa}\ \emph {et~al.}(2010)\citenamefont
  {Vilarrasa}, \citenamefont {Bolster}, \citenamefont {Dentz}, \citenamefont
  {Olivella},\ and\ \citenamefont {Carrera}}]{vilarrasa2010comp}%
  \BibitemOpen
  \bibfield  {author} {\bibinfo {author} {\bibfnamefont {Victor}\ \bibnamefont
  {Vilarrasa}}, \bibinfo {author} {\bibfnamefont {Diogo}\ \bibnamefont
  {Bolster}}, \bibinfo {author} {\bibfnamefont {Marco}\ \bibnamefont {Dentz}},
  \bibinfo {author} {\bibfnamefont {Sebastia}\ \bibnamefont {Olivella}}, \ and\
  \bibinfo {author} {\bibfnamefont {Jesus}\ \bibnamefont {Carrera}},\
  }\bibfield  {title} {\enquote {\bibinfo {title} {Effects of {CO}$_2$
  compressibility on {CO}$_2$ storage in deep saline aquifers},}\ }\href@noop
  {} {\bibfield  {journal} {\bibinfo  {journal} {Transport in porous media}\
  }\textbf {\bibinfo {volume} {85}},\ \bibinfo {pages} {619--639} (\bibinfo
  {year} {2010})}\BibitemShut {NoStop}%
\bibitem [{\citenamefont {Teige}\ \emph {et~al.}(2006)\citenamefont {Teige},
  \citenamefont {Thomas}, \citenamefont {Hermanrud}, \citenamefont {{\O}ren},
  \citenamefont {Rennan}, \citenamefont {Wilson},\ and\ \citenamefont
  {{Nordg{\r{a}}rd Bol{\r{a}}s}}}]{teige-jgr-2006}%
  \BibitemOpen
  \bibfield  {author} {\bibinfo {author} {\bibfnamefont {G.~M.~G.}\
  \bibnamefont {Teige}}, \bibinfo {author} {\bibfnamefont {W.~L.~H.}\
  \bibnamefont {Thomas}}, \bibinfo {author} {\bibfnamefont {C.}~\bibnamefont
  {Hermanrud}}, \bibinfo {author} {\bibfnamefont {P.-E.}\ \bibnamefont
  {{\O}ren}}, \bibinfo {author} {\bibfnamefont {L.}~\bibnamefont {Rennan}},
  \bibinfo {author} {\bibfnamefont {O.~B.}\ \bibnamefont {Wilson}}, \ and\
  \bibinfo {author} {\bibfnamefont {H.~G.}\ \bibnamefont {{Nordg{\r{a}}rd
  Bol{\r{a}}s}}},\ }\bibfield  {title} {\enquote {\bibinfo {title} {Relative
  permeability to wetting-phase water in oil reservoirs},}\ }\href@noop {}
  {\bibfield  {journal} {\bibinfo  {journal} {Journal of Geophysical Research}\
  }\textbf {\bibinfo {volume} {111}},\ \bibinfo {pages} {B12204} (\bibinfo
  {year} {2006})}\BibitemShut {NoStop}%
\bibitem [{\citenamefont {MacMinn}\ and\ \citenamefont
  {Juanes}(2009)}]{macminn-compgeosci-2009}%
  \BibitemOpen
  \bibfield  {author} {\bibinfo {author} {\bibfnamefont {C.~W.}\ \bibnamefont
  {MacMinn}}\ and\ \bibinfo {author} {\bibfnamefont {R.}~\bibnamefont
  {Juanes}},\ }\bibfield  {title} {\enquote {\bibinfo {title} {Post-injection
  spreading and trapping of {CO}$_2$ in saline aquifers: Impact of the plume
  shape at the end of injection},}\ }\href@noop {} {\bibfield  {journal}
  {\bibinfo  {journal} {Computational Geosciences}\ }\textbf {\bibinfo {volume}
  {13}},\ \bibinfo {pages} {483--491} (\bibinfo {year} {2009})}\BibitemShut
  {NoStop}%
\bibitem [{\citenamefont {MacMinn}\ \emph {et~al.}(2011)\citenamefont
  {MacMinn}, \citenamefont {Szulczewski},\ and\ \citenamefont
  {Juanes}}]{macminn-jfm-2011}%
  \BibitemOpen
  \bibfield  {author} {\bibinfo {author} {\bibfnamefont {C.~W.}\ \bibnamefont
  {MacMinn}}, \bibinfo {author} {\bibfnamefont {M.~L.}\ \bibnamefont
  {Szulczewski}}, \ and\ \bibinfo {author} {\bibfnamefont {R.}~\bibnamefont
  {Juanes}},\ }\bibfield  {title} {\enquote {\bibinfo {title} {{CO}$_2$
  migration in saline aquifers. {P}art 2. {C}apillary and solubility
  trapping},}\ }\href@noop {} {\bibfield  {journal} {\bibinfo  {journal}
  {Journal of Fluid Mechanics}\ }\textbf {\bibinfo {volume} {688}},\ \bibinfo
  {pages} {321--351} (\bibinfo {year} {2011})}\BibitemShut {NoStop}%
\end{thebibliography}
%

\end{document}